\PassOptionsToPackage{pdfpagelabels=false}{hyperref}
\documentclass[useAMS,usenatbib]{mnras}
\pdfoutput=1
\usepackage[pdftex]{graphicx}
\usepackage{url}
\usepackage{amsmath}
\usepackage{natbib}
\usepackage{geometry}
\usepackage{txfonts}
\usepackage{times}
\usepackage{multirow}

\def \MSUN{{\rm M}_{\odot}}

\def \arrvline{\hfil\kern0pt\vline\kern0pt\hfilneg}

\title[Introducing Project GIBLE]{Zooming in on the circumgalactic medium: resolving small-scale gas structure with the GIBLE cosmological simulations}
\author[R. Ramesh \& D. Nelson]{Rahul Ramesh$^{1}$\thanks{E-mail: rahul.ramesh@stud.uni-heidelberg.de} and Dylan Nelson$^{1}$\\
$^{1}$ Universität Heidelberg, Zentrum für Astronomie, Institut für theoretische Astrophysik, Albert-Ueberle-Str. 2, 69120 Heidelberg, Germany\\
}

\date{}
\pubyear{2023}

\begin{document}

\maketitle

\begin{abstract}
We introduce Project GIBLE (Gas Is Better resoLved around galaxiEs), a suite of cosmological zoom-in simulations where gas in the circumgalactic medium (CGM) is preferentially simulated at ultra-high numerical resolution. Our initial sample consists of eight galaxies, all selected as Milky Way-like galaxies at $z=0$ from the TNG50 simulation. Using the same galaxy formation model as IllustrisTNG, and the moving-mesh code \textsc{arepo}, we re-simulate each of these eight galaxies maintaining a resolution equivalent to TNG50-2 ($m_{\rm{gas}}$\,$\sim$\,$8 \times 10^5~\MSUN$). However, we use our super-Lagrangian refinement scheme to more finely resolve gas in the CGM around these galaxies. Our highest resolution runs achieve 512 times better mass resolution ($\sim$\,$10^3~\MSUN$). This corresponds to a median spatial resolution of $\sim$\,$75$~pc at $0.15~R_{\rm{200,c}}$, which coarsens with increasing distance to $\sim$\,$700$~pc at the virial radius. We make predictions for the covering fractions of several observational tracers of multi-phase CGM gas: HI, MgII, CIV and OVII. We then study the impact of improved resolution on small scale structure. While the abundance of the smallest cold, dense gas clouds continues to increase with improving resolution, the number of massive clouds is well converged. We conclude by quantifying small scale structure with the velocity structure function and the auto-correlation function of the density field, assessing their resolution dependence. The GIBLE cosmological hydrodynamical simulations enable us to improve resolution in a computationally efficient manner, thereby achieving numerical convergence of a subset of key CGM gas properties and observables.
\end{abstract}


\section{Introduction}

Galaxy formation and evolution is a complex process. Over the years, various cosmological simulation projects have attempted to improve our understanding of the physics governing cosmic growth. While the first set of cosmological simulations were dark matter-only, more recent simulations over the past decade or so have begun including important baryonic processes, thereby making it possible to statistically reproduce baryonic structures of the observed universe \citep[for a recent overview of cosmological simulations, see][]{vogelsberger2020}.

These cosmological hydrodynamical simulations have often fared well when reproducing galaxy- and halo-scale global properties. For instance, the stellar size-mass relation obtained from simulations \citep[e.g.][]{genel2018,henden2020,feldmann2023} is increasingly consistent with those of observations \citep[e.g.][]{shen2003,bernardi2014}.
Simulated galaxies can separate into two populations based on their star formation activity, thereby giving rise to a bimodal galaxy color distribution \citep[e.g.][]{nelson2018,cui2021}, akin to observational studies \citep[e.g][]{strateva2001,balogh2004}. Quenched fractions of simulated satellite galaxies are found to correlate with their environment \citep[e.g.][]{wright2019,donnari2021}, as shown by observations \cite[e.g.][]{geha2012,medling2018}.

However, the consistency between cosmological hydrodynamical simulations and observations is less straightforward with respect to the distribution of gas around galaxies -- the circumgalactic medium (CGM). This reservoir of gas is believed to be multi-scale and multi-phase, with cold, dense clouds embedded in a volume filling warm-hot phase, making it a challenge to numerically simulate (see \citealt{donahue2022} for a recent review of the CGM).

Cosmological hydrodynamical simulations typically make robust predictions for the properties of CGM gas on relatively larger scales and lower densities.
For instance, the warm-hot phase, traced by OVI, OVII and OVIII, is believed to be decently (numerically) converged at typical resolutions realised by modern cosmological simulations \citep[e.g.][]{nelson2018b,wijers2019}, returning similar column density distribution functions as observations \citep[e.g.][]{thom2008,danforth2016}. At the Milky Way mass, the IllustrisTNG simulations produce AGN feedback-driven bubbles of hot, rarified gas on either side of galactic disks \citep{pillepich2021}, much like the Fermi/eROSITA bubbles emerging from the galactic centre of the Milky Way \citep{su2010, predehl2020}. Observed angular anisotropies of metallicity, density and magnetic field strengths in CGM gas \citep{cameron2021,zhang2022,heesen2023} are also found in some simulations \citep[][]{peroux2020,vandevoort2021,truong2021,ramesh2023c}. 

However, the picture differs for the more clumpy, colder, dense phase. Some studies, although not all, find higher HI covering fractions with improving resolution, without any clear sign of convergence \citep{peeples2019,voort2019}. Recent work also suggests that improving numerical resolution results in the existence of a larger number of cold, dense discrete CGM gas clouds \citep{nelson2020, ramesh2023b}. In reality, observations of clouds in the Milky Way \citep[e.g.][]{hsu2011} and extragalactic halos \citep[e.g.][]{zahedy2019} reveal the existence of cloud sizes at least as small as $\sim$\,$100-200$~pc, thereby requiring simulations to adequately sample the underlying gas distribution on these scales to capture such objects. Whether or not resolving the small scale structure in the CGM has a qualitative or quantitative impact on galaxy formation remains an open question.

While recent cosmological large-volume and zoom-in simulations are significantly better resolved than those of the past \citep[see e.g.][]{pillepich2023}, it is still a challenge to simulate CGM gas at the level of resolution discussed above. As a result, various idealised simulations are commonly used to study phenomena at these scales \citep[e.g.][]{scannapieco2015,goldsmith2016,gronke2020}. Such numerical experiments can explore small-scale physics, although at the expense of lacking the realism of a cosmological CGM.

Recently, an alternative approach has emerged, using cosmological hydrodynamical simulations that preferentially increase the resolution of CGM gas while maintaining a relatively coarse resolution in other regions, most importantly the galaxy itself \citep{hummels2019,peeples2019,suresh2019,voort2019}. This greatly reduces the cost with respect to standard simulations, as computational effort is diverted from dense ISM gas in the galaxy, to the CGM. These simulations provide the ideal bridge between highly resolved idealised setups, and cosmological setups that naturally capture the complexity of the CGM.

In this paper, we describe and introduce a new set of simulations run with our `CGM refinement' technique, an improvement of past work \citep{suresh2019}.
These are the GIBLE (Gas Is Better resoLved around galaxiEs) simulations, currently comprised of eight Milky Way-like galaxies at $z=0$, drawn from the TNG50 simulation. The rest of this paper is organised as follows: in Section~\ref{methods}, we describe the galaxy formation model and the CGM refinement technique used. Results are presented in Section~\ref{results}, and we summarise the paper in Section~\ref{summary}. 

\begin{figure*}
\centering 
\includegraphics[width=14cm]{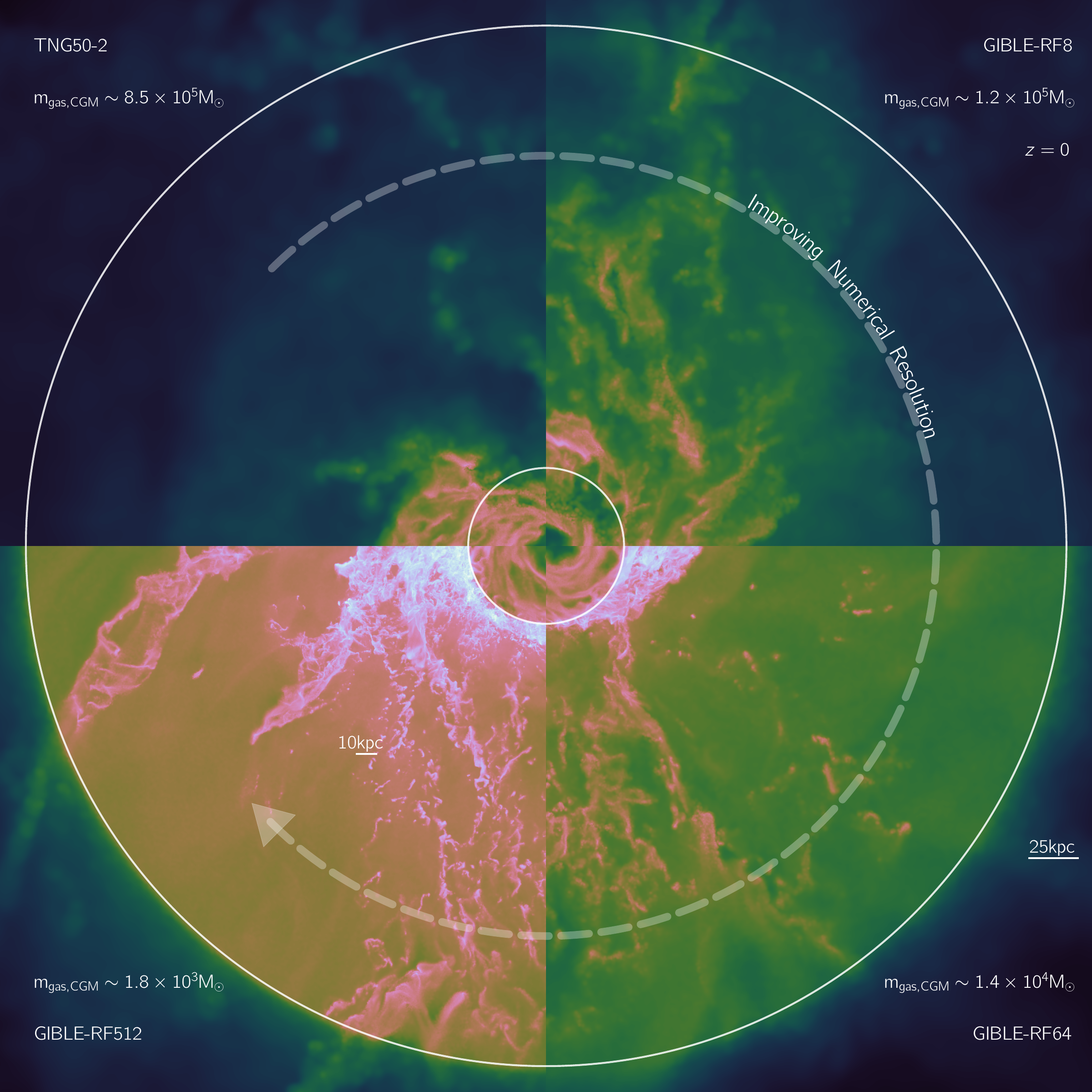}
\includegraphics[width=6cm]{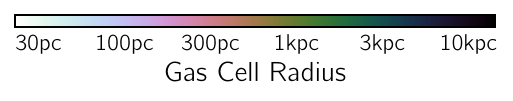}
\caption{A qualitative illustration of our refinement scheme. Each quadrant shows a part of the same halo from four different simulations. The top left corresponds to TNG50-2, while the others are taken from GIBLE. Starting from the top-left, the (CGM gas) resolution increases in the clockwise direction. The image extends $\pm~1.05~R_{\rm{200,c}}$ along the plane, and $\pm~20$~kpc perpendicular to the plane. Colors show the gas cell radius, i.e. the spatial resolution, with the lighter pixels corresponding to better numerical resolution. At higher resolutions, finer structure is apparent on smaller scales, embedded in an altogether better resolved halo.}
\label{fig:mainVisFigure}
\end{figure*}

\section{Methods}\label{methods}

\subsection{Galaxy Formation Model and Initial Conditions}

Project GIBLE uses the well-studied and validated IllustrisTNG model \citep{weinberger2017, pillepich2018} to simulate the basic elements of galaxy formation and evolution, within the moving mesh code \textsc{arepo} \citep{springel2010}.
The model accounts for star formation, stellar and AGN feedback, stellar enrichment, metal-line cooling of gas, and other important processes, along with the inclusion of magnetic fields \citep{pakmor2014}. The star formation and stellar feedback model is the subgrid two-phase model of \cite{springel2003}.

The IllustrisTNG project is made of three suites of simulations: the relatively large volume boxes of TNG100 and TNG300 \citep{naiman2018, nelson2018, pillepich2018b, marinacci2018, springel2018}, and the high-resolution, smaller box TNG50 \citep{nelson2019, pillepich2019}. Project GIBLE (re-)simulates eight galaxies from the TNG50 simulation, all of which are a part of the Milky Way-like (MW-like) sample from \cite{ramesh2023a}\footnote{This sample is a subset of the MW/M31-like galaxies of TNG50 presented in \citealt{pillepich2023}.}. To recap, these are a set of galaxies that (i) have a stellar mass in the range $10^{10.5}\,\rm{M}_\odot$ to $10^{10.9}\,\rm{M}_\odot$, measured within a 3D aperture of 30kpc, (ii) have no other galaxy having M$_\star > 10^{10.5}~\rm{M}_\odot$ within a distance of $500$\,kpc, (iii) reside in halos that satisfy virial mass, M$_{\rm{200,c}} < 10^{13}~\rm{M}_\odot$, (iv) are disky, either based on a constraint on the minor-to-major axis ratio of the stellar mass distribution ($s < 0.45$) or through visual inspection of stellar light maps, and (v) lie at the potential minimum of their host halos, as identified by the friends-of-friends algorithm \citep{davies1985}, i.e. they are central galaxies.

While the MW-like sample is assembled using TNG50-1, the highest resolution version of the TNG50 box, we derive initial conditions for Project GIBLE from TNG50-2 (lower resolution by a factor eight in mass), following a cross-match of galaxies across these two runs. The eight galaxies in our sample have been carefully selected to ensure a good amount of diversity: two have been selected based on the presence of an SMC- and LMC-like satellite within a distance of 300kpc \citep{engler2021}, two that host Fermi-/eROSITA-like bubbles on either side of the central galaxy \citep{pillepich2021}, two that possess a bar and/or stellar disk length similar to the central galaxy, one with a M31-like companion \citep[i.e a galaxy with a similar mass as M31 that is moving towards the central MW-like galaxy;][]{pillepich2023}, and the eighth is chosen at random from the aforementioned set of 132 TNG50 MW-like galaxies. The galaxy with the M31-like companion does not have a significant cold gas disk in TNG50-2 at $z=0$, due to prior disruption by AGN feedback during the quenching process. This galaxy is an outlier in terms of multiple properties, adding to the diversity of the sample.

With the sample selected, we construct zoom initial conditions using the multi-mass version of the \textsc{n-genic} code \citep{angulo2012,springel2021}. At $z=0$, we identify particles that are within 4 times the halo virial radius $R_{\rm{200c}}$. The high resolution region is then constructed using the convex hull of the $z=127$ positions of these particles. The rest of the box is populated with resolution elements of higher mass, with a gradual worsening of resolution with increasing distance from the central zoom region.

\subsection{Super-Lagrangian Refinement Scheme}

Project GIBLE uses an improvement of the `CGM refinement' previously explored in \cite{suresh2019}, with several technical updates to the underlying code. The most major change concerns the functionality of black hole particles: earlier, it was possible to seed, i.e. create, at most a single black hole particle. Moreover, this black hole particle was only a psuedo-particle, with accretion and subsequent feedback processes disabled. Our new CGM refinement technique removes these limitations, and black holes now function as normal TNG black holes \citep{weinberger2017}. 

Black holes are important in our refinement scheme as they are used to trace the centre of the target halo, and thus the refinement region where CGM gas is super-refined \citep{suresh2019}. Since only one refinement region is generally desired per run, the earlier CGM Zoom refinement scheme limited the number of seeded black holes to one. However, to be consistent with the TNG runs and the black hole feedback models, we now allow for the possibility of multiple black holes to be seeded, but `flag' the first one to be formed. By default, this takes place in the first halo to cross a threshold halo mass of $\sim$\,$7 \times 10^{10}~\rm{M_\odot}$. When the flagged black holes merges, the flag is passed on to the descendent, thus ensuring that the flagged black hole is never lost through the course of the simulation.

\begin{table}
\centering
\begin{tabular}{||c|c|c|c||} 
 \hline
 Refinement & $m_{\rm{gas, galaxy}}$ & $m_{\rm{gas, CGM}}$ & Compute Time\\ 
Factor & $[\rm{M_\odot}]$ & $[\rm{M_\odot}]$ & [$10^6$ CPU Hrs] \\ [0.5ex] 
 \hline\hline
 8 & $\sim 8.5 \times 10^5$ & $\sim 1.2 \times 10^5$ & $\sim 0.05$ \\ 
 64 & $\sim 8.5 \times 10^5$ & $\sim 1.4 \times 10^4$ & $\sim 0.1$ \\
 512 & $\sim 8.5 \times 10^5$ & $\sim 1.8 \times 10^3$ & $\sim 0.5$ \\ [1ex] 
 \hline
\end{tabular}
\caption{Characteristic numbers associated with our simulations: the first column lists the refinement factor (RF), i.e. the factor by which CGM gas is better resolved compared to the galaxy. The second column specifies the average gas mass resolution in the galaxy, which is roughly the same for all runs. The average CGM gas mass resolution is shown in the third column, which progressively improves from top to bottom. The average computational time for our sample of MW-like halos run to $z=0$ is shown in the last column.}
\label{table:global_numbers}
\end{table}

\begin{figure*}
\centering 
\includegraphics[width=8cm]{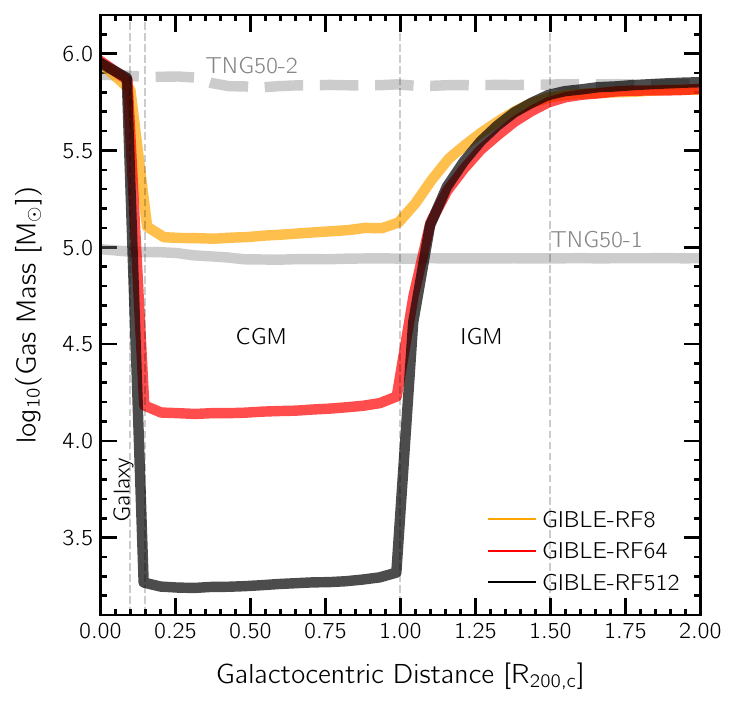}
\includegraphics[width=8.22cm]{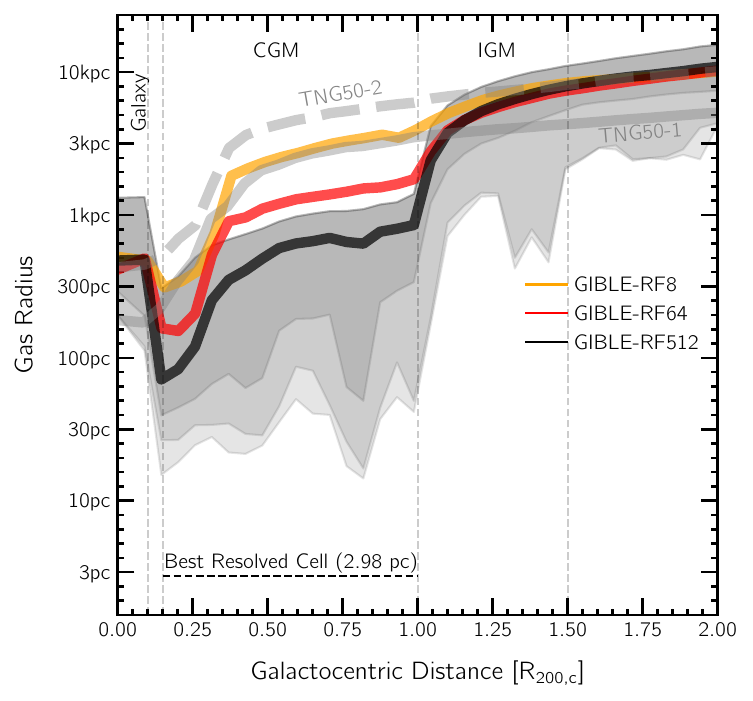}
\caption{A schematic illustration of our refinement scheme. Orange, red and black curves correspond to RF8, RF64 and RF512 respectively, while the two gray lines correspond to TNG50. We show the radial profile of mass resolution on the left, and spatial resolution on the right. The different dashed vertical lines signify the different boundaries used to define the refinement region. The distance dependent target mass of our CGM refinement scheme is used to achieve a boosted mass/spatial resolution in the CGM region.}
\label{fig:mainQuanFigure}
\end{figure*}

Our refinement scheme builds on the already existing (de)refinement operations that \textsc{arepo} performs to maintain (high-resolution) gas cells within a factor of two\footnote{Note that, in principle, this factor can be set arbitrarily. A value of two is employed in the TNG runs.} of a given (constant) \textit{target mass}, $m_{\rm{target}}$ \citep{springel2010}. Gas cells whose mass falls below the defined threshold, i.e $< 0.5 \times m_{\rm{target}}$, are dissolved and their properties are conservatively distributed to their Voronoi neighbors. Those whose mass rises above the upper limit, i.e $> 2.0 \times m_{\rm{target}}$, are split into two. This procedure ensures that all (high-resolution) gas cells contain roughly the same mass, at least up to a desired threshold. In our scheme, the constant target mass is replaced by a \textit{distance dependent target mass}, $m_{\rm{dist, target}}(r)$, which is defined for every high-resolution gas cell, based on its distance from the flagged black hole, $r$:

\begin{equation}
    m_{\rm{dist, target}}(r) = \begin{cases}
      m_{\rm{target}} & \text{if $r< r_{\rm{galaxy}}$}\\
      m_{\rm{target}} / \text{RF} & \text{if $ r_{\rm{CGM,min}}<r< r_{\rm{CGM,max}}$}\\
      m_{\rm{target}} & \text{if $r> r_{\rm{IGM}}$}
    \end{cases}
\label{eq:target_gas_mass}    
\end{equation}

where $r_{\rm{galaxy}}$, $r_{\rm{CGM,min}}$, $r_{\rm{CGM,max}}$ and $r_{\rm{IGM}}$ are the radii of the galaxy, inner boundary of the CGM, outer boundary of the CGM, and the IGM, respectively. The variable RF is the `refinement factor', by which CGM gas is better refined with respect to $m_{\rm{target}}$, and hence better refined with respect to gas in the galaxy. These five values remain constant throughout the run. To avoid resolution jumps, at intermediate buffer regions between (a) the galaxy and the inner boundary of the CGM, and (b) the outer boundary of the CGM and the IGM, the target mass is (linearly) interpolated.

\begin{figure}
\centering 
\includegraphics[width=8cm]{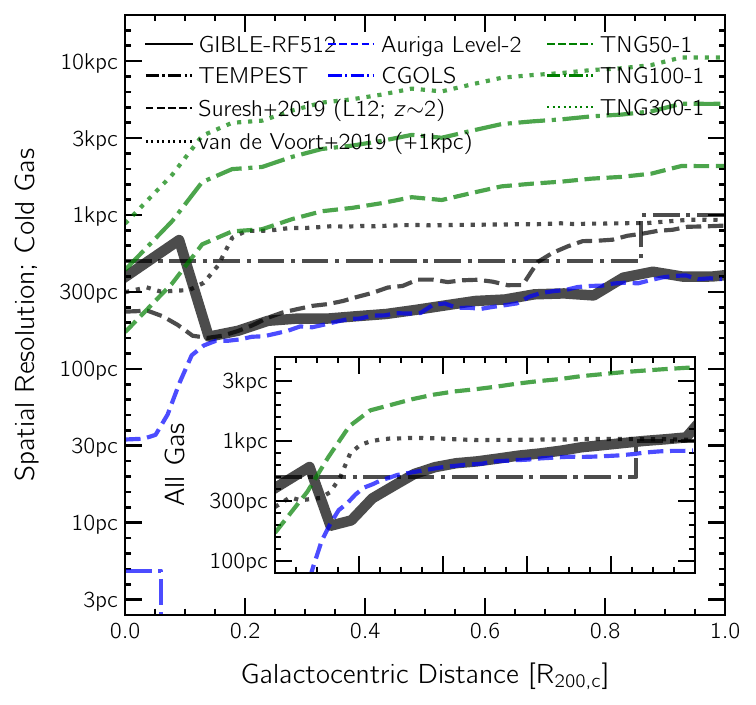}
\caption{A comparison between our simulation(s) and other projects through radial profiles of spatial resolution. The three green curves correspond to the three TNG boxes at $z=0$, i.e. large volume uniform-resolution simulations. The blue curves are examples of projects that simulate a smaller volume at high resolution, either in a cosmological (Auriga; dashed) or an idealised (CGOLS; dot-dashed) context, but with uniform resolution, either mass or spatial, within the region of interest. The black curves show those that employ a CGM-refinement technique, all run as cosmological zoom-in simulations, with a further improvement in resolution within a refinement region that traces the CGM. Preferentially increasing mass resolution has a greater impact on the cold phase (as in the case of GIBLE and \protect\citealt{suresh2019}), while improved spatial resolution has a more substantial effect on the hot phase (as is the case with the other techniques).}
\label{fig:comp_projects}
\end{figure}

In all our runs, we define [$r_{\rm{galaxy}}$, $r_{\rm{CGM,min}}$, $r_{\rm{CGM,max}}$, $r_{\rm{IGM}}$] = [0.10, 0.15, 1.00, 1.50] $\times~R_{\rm{200,c}}$, where $R_{\rm{200,c}}$ is the $z=0$ virial radius of the corresponding halo from TNG50-2. As a result, a constant comoving volume bounded by $r_{\rm{CGM,min}}$ and $r_{\rm{CGM,max}}$ is maintained at super-high resolution (i.e. a factor of RF better than $m_{\rm{target}}$) between the point the first black hole is seeded ($z$\,$\sim$\,$3-5$ for the halos in our sample) and $z=0$. At all redshifts, we thus always define the CGM as the region between these two (pre-specified) radii.

We provide a visual illustration of our technique in Figure~\ref{fig:mainVisFigure}. Each quadrant corresponds to a different resolution level of a given galaxy. The top-left corresponds to the halo from TNG50-2 (CGM gas mass resolution, $m_{\rm{gas, CGM}}$\,$\sim$\,$8.5 \times 10^5 \rm{M_\odot}$\footnote{In TNG50-2, this resolution is maintained throughout the box.}), the top right to the RF8 run of the corresponding halo of GIBLE ($m_{\rm{gas, CGM}}$\,$\sim$\,$1.2\times 10^5 \rm{M_\odot}$), the bottom right to RF64 ($m_{\rm{gas, CGM}}$\,$\sim$\,$1.4\times 10^4 \rm{M_\odot}$), and the bottom-left to RF512 ($m_{\rm{gas, CGM}}$\,$\sim$\,$1.8\times 10^3 \rm{M_\odot}$). Background colors show the average gas cell radius of gas in a $\pm 20$~kpc depth projection. The two circles are drawn at radii [0.15, 1.00] $\times R_{\rm{200,c}}$, the two boundaries of the CGM, and the circular arrow signifies the direction of improving numerical resolution, i.e. clockwise from the top-left to bottom-left. While the background CGM, on average, shifts towards brighter colors along this arrow, signifying better numerical resolution throughout the halo, a greater amount of small scale structure is also apparent. The RF512 run resolves the denser portions of the CGM with cell radii of 10s of parsecs.

A more quantitative illustration of the scheme is provided in Figure~\ref{fig:mainQuanFigure}. Both panels show trends of numerical resolution as a function of galactocentric distance normalised by the virial radius. Curves from RF8, RF64 and RF512 are shown through orange, red and black curves, respectively. The two gray curves correspond to TNG50-1 (solid) and TNG50-2 (dashed). The vertical lines portray the definitions we use for the radii of the galaxy, the two boundaries of the CGM, and the IGM.

The left panel shows the mass resolution as a function of distance, which traces Equation~\ref{eq:target_gas_mass}. By definition, at distances $< 0.1~R_{\rm{200,c}}$, i.e. within the galaxy, the mass resolution of all three resolution runs of GIBLE are roughly the same, and roughly equal to the resolution of TNG50-2.\footnote{This resolution, i.e. TNG100-1, is where the TNG model is calibrated versus observational constraints \citep{weinberger2017,pillepich2018}, and hence the regime where the models for galactic feedback return the most realistic and robust results. That is, the energetics across this `inner boundary condition' from the galaxy itself into the CGM are best when we maintain the galaxy at TNG50-2 resolution, as in our simulations.} The same holds for distances $> 1.5~R_{\rm{200,c}}$, i.e. beyond the IGM. Between $0.15~R_{\rm{200,c}}$ and $R_{\rm{200,c}}$, i.e. in the CGM, resolution of the RF8 run is similar to TNG50-1, although not exactly the same, since we only (de)refine gas cells if they are off by a factor of two with respect to the (distance dependent) target gas mass. The RF64 curve (dashed) is vertically offset by a factor of eight, and the RF512 curve (solid) further by another factor of eight. The three GIBLE curves converge in the two buffer regions, eventually overlapping each other in the galaxy and beyond the IGM. The two TNG50 curves show constant mass resolution at all radii, which is also what would be seen in a `normal' zoom-in simulation, at least within the high-resolution region. Table~\ref{table:global_numbers} succinctly summarises the various resolution levels realised by our three RF levels. 

The right panel shows how spatial resolution changes with distance, for one of our eight simulated halos. Similar to the left panel, the three GIBLE curves converge with TNG50-2 in the galaxy, and beyond the IGM. This corresponds to a median gas cell size of $\sim$\,$450$~pc in the galaxy, and $\gtrsim$\,$10$~kpc in the far reaches of the IGM. In the CGM, the spatial resolution progressively increases for the higher RF-runs: the RF8 curve (orange) is in the ballpark of TNG50-1 (solid gray), with a median cell size of $\sim$\,$300$~pc at $0.15~R_{\rm{200,c}}$, increasing gradually to $\sim$\,$3$~kpc at the virial radius. The red (RF64) and black (RF512) curves are better resolved by factors of $\sim$ 2 and 4, respectively, at all distances, resulting in a substantially improved median spatial resolution of $\sim$\,$75$~pc at $0.15~R_{\rm{200,c}}$ for the RF512 run.

The three shaded regions show three different percentile regions for the RF512 run: $5^{\rm{th}}$-$95^{\rm{th}}$, $0.1^{\rm{th}}$-$95^{\rm{th}}$ and $0.0001^{\rm{th}}$-$95^{\rm{th}}$ for the darkest to lightest bands, i.e. the latter two bands portray the extreme outliers of the distribution. The $5^{\rm{th}}$-$95^{\rm{th}}$ region is relatively broad, as a result of multi-phase gas at all distances. The dips in the percentile regions seen at, e.g. $\sim$\,$0.8~R_{\rm{200,c}}$, are a result of ISM gas of satellites being super-refined when they pass through the refinement region.
The best resolved cell of this RF512 run is $\sim$\,$3$~pc ($n_H$\,$\sim$\,$2000~\rm{cm^{-3}}$), located in the dense ISM of an infalling satellite, and is shown by the dashed horizontal line at the bottom end of the panel. This is roughly $4$ times smaller than the best resolved gas cell ($\sim$\,$10$~pc) of the TNG50 box \citep{pillepich2019}.

\begin{table*}
\centering
\begin{tabular}{||c|c|c|c|c|c|c|c|c|c|c|c|c|c|c|c||} 
 \hline
 H\# & RF & \arrvline & ${\rm{M_{200,c}}}$ & $\rm{R_{200,c}}$ & $\rm{M_{\star}}$ & $\rm{M_{BH}}$ & N$_{\rm{merger}, z<2}$ & \arrvline & $\rm{M_{g, CGM}}$ & f$_{\rm{m, CGM}}$ & N$_{\rm{cld, CGM}}$ & $\rm{\overline{T_{CGM}}}$ &f$_{\rm{vol, CGM}}$ [\%]\\ 
   & & \arrvline & [log $ \rm{M_\odot}]$ & [kpc] & [log $ \rm{M_\odot}]$ & [log $ \rm{M_\odot}]$ & Ma.,~Mi.,~A & \arrvline & [log $ \rm{M_\odot}]$ & [C; \%] &  & [log K] & C,~W,~H \\ [0.05ex] 

 \hline\hline
 \multirow{3}{*}{S91} & \multicolumn{1}{c}{8} & \arrvline & \multicolumn{1}{c}{12.18} & \multicolumn{1}{c}{242.74} &
 \multicolumn{1}{c}{10.57} &
 \multicolumn{1}{c}{8.40} &
 \multicolumn{1}{c}{0,~0,~16} & \arrvline &
 \multicolumn{1}{c}{10.12} & \multicolumn{1}{c}{0.75} & \multicolumn{1}{c}{18} & \multicolumn{1}{c}{5.95} & \multicolumn{1}{c}{00.19,~18.37,~81.44} \\\cline{2-14} 
 & \multicolumn{1}{c}{64} & \arrvline & \multicolumn{1}{c}{12.18} & \multicolumn{1}{c}{243.23} & 
 \multicolumn{1}{c}{10.61} &
 \multicolumn{1}{c}{8.33} &
 \multicolumn{1}{c}{0,~0,~18} & \arrvline &
 \multicolumn{1}{c}{10.31} &  \multicolumn{1}{c}{0.07} & \multicolumn{1}{c}{15} & \multicolumn{1}{c}{5.90} & \multicolumn{1}{c}{00.10,~95.60,~04.30}\\\cline{2-14}
 & \multicolumn{1}{c}{512} & \arrvline & \multicolumn{1}{c}{12.19} & \multicolumn{1}{c}{243.28} & 
 \multicolumn{1}{c}{10.53} &
 \multicolumn{1}{c}{8.32} &
 \multicolumn{1}{c}{0,~0,~21} & \arrvline &
 \multicolumn{1}{c}{10.05} & \multicolumn{1}{c}{0.59} & \multicolumn{1}{c}{201} & \multicolumn{1}{c}{5.87} & \multicolumn{1}{c}{00.24,~29.19,~70.56} \\\hline\hline

 \multirow{3}{*}{S98} & \multicolumn{1}{c}{8} & \arrvline & \multicolumn{1}{c}{12.31} & \multicolumn{1}{c}{266.84} &
 \multicolumn{1}{c}{10.72} &
 \multicolumn{1}{c}{8.33} &
 \multicolumn{1}{c}{2,~0,~15} & \arrvline & \multicolumn{1}{c}{11.13} & \multicolumn{1}{c}{47.51} & \multicolumn{1}{c}{1175} & \multicolumn{1}{c}{4.80} & \multicolumn{1}{c}{12.44,~82.29,~05.26} \\\cline{2-14} 
 & \multicolumn{1}{c}{64} & \arrvline & \multicolumn{1}{c}{12.32} & \multicolumn{1}{c}{270.62} & 
 \multicolumn{1}{c}{10.82} &
 \multicolumn{1}{c}{8.30} &
 \multicolumn{1}{c}{1,~0,~15} & \arrvline & \multicolumn{1}{c}{11.20} & \multicolumn{1}{c}{51.24} & \multicolumn{1}{c}{8918} & \multicolumn{1}{c}{4.79} & \multicolumn{1}{c}{13.10,~69.59,~17.31} \\\cline{2-14}
 & \multicolumn{1}{c}{512} & \arrvline & \multicolumn{1}{c}{12.33} & \multicolumn{1}{c}{271.19} & 
 \multicolumn{1}{c}{10.84} &
 \multicolumn{1}{c}{8.28} &
 \multicolumn{1}{c}{1,~0,~17} & \arrvline & \multicolumn{1}{c}{11.20} & \multicolumn{1}{c}{47.60} & \multicolumn{1}{c}{61319} & \multicolumn{1}{c}{4.86} & \multicolumn{1}{c}{14.11,~76.80,~09.09} \\\hline\hline

 \multirow{3}{*}{S105} & \multicolumn{1}{c}{8} & \arrvline & \multicolumn{1}{c}{12.28} & \multicolumn{1}{c}{261.91} &
 \multicolumn{1}{c}{10.63} &
 \multicolumn{1}{c}{8.13} &
 \multicolumn{1}{c}{1,~2,~09} & \arrvline & \multicolumn{1}{c}{11.08} & \multicolumn{1}{c}{16.79} & \multicolumn{1}{c}{1844} & \multicolumn{1}{c}{5.29} & \multicolumn{1}{c}{06.50,~85.02,~08.48} \\\cline{2-14} 
 & \multicolumn{1}{c}{64} & \arrvline & \multicolumn{1}{c}{12.29} & \multicolumn{1}{c}{263.46} & 
 \multicolumn{1}{c}{10.65} &
 \multicolumn{1}{c}{8.07} &
 \multicolumn{1}{c}{1,~0,~14} & \arrvline & \multicolumn{1}{c}{11.12} & \multicolumn{1}{c}{16.79} & \multicolumn{1}{c}{9658} & \multicolumn{1}{c}{5.34} & \multicolumn{1}{c}{04.47,~95.16,~00.37} \\\cline{2-14}
 & \multicolumn{1}{c}{512} & \arrvline & \multicolumn{1}{c}{12.29} & \multicolumn{1}{c}{263.34} & 
 \multicolumn{1}{c}{10.65} &
 \multicolumn{1}{c}{8.10} &
 \multicolumn{1}{c}{2,~3,~07} & \arrvline & \multicolumn{1}{c}{11.12} & \multicolumn{1}{c}{22.30} & \multicolumn{1}{c}{40916} & \multicolumn{1}{c}{5.22} & \multicolumn{1}{c}{05.32,~91.58,~03.10} \\\hline\hline

 \multirow{3}{*}{S146} & \multicolumn{1}{c}{8} & \arrvline & \multicolumn{1}{c}{12.19} & \multicolumn{1}{c}{245.03} &
 \multicolumn{1}{c}{10.74} &
 \multicolumn{1}{c}{8.11} &
 \multicolumn{1}{c}{0,~0,~07} & \arrvline & \multicolumn{1}{c}{11.08} & \multicolumn{1}{c}{32.03} & \multicolumn{1}{c}{1414} & \multicolumn{1}{c}{5.13} & \multicolumn{1}{c}{05.82,~85.62,~08.57} \\\cline{2-14} 
 & \multicolumn{1}{c}{64} & \arrvline & \multicolumn{1}{c}{12.16} & \multicolumn{1}{c}{239.48} & 
 \multicolumn{1}{c}{10.74} &
 \multicolumn{1}{c}{8.17} &
 \multicolumn{1}{c}{0,~0,~08} & \arrvline & \multicolumn{1}{c}{10.81} & \multicolumn{1}{c}{37.17} & \multicolumn{1}{c}{2784} & \multicolumn{1}{c}{5.21} & \multicolumn{1}{c}{01.93,~23.76,~74.31} \\\cline{2-14}
 & \multicolumn{1}{c}{512} & \arrvline & \multicolumn{1}{c}{12.21} & \multicolumn{1}{c}{247.74} & 
 \multicolumn{1}{c}{10.78} &
 \multicolumn{1}{c}{8.09} &
 \multicolumn{1}{c}{0,~0,~15} & \arrvline & \multicolumn{1}{c}{11.11} & \multicolumn{1}{c}{40.13} & \multicolumn{1}{c}{39448} & \multicolumn{1}{c}{4.97} & \multicolumn{1}{c}{07.84,~91.21,~00.95} \\\hline\hline

 \multirow{3}{*}{S167} & \multicolumn{1}{c}{8} & \arrvline & \multicolumn{1}{c}{12.17} & \multicolumn{1}{c}{242.12} &
 \multicolumn{1}{c}{10.85} &
 \multicolumn{1}{c}{8.17} &
 \multicolumn{1}{c}{0,~1,~20} & \arrvline & \multicolumn{1}{c}{10.89} & \multicolumn{1}{c}{8.39} & \multicolumn{1}{c}{1238} & \multicolumn{1}{c}{5.53} & \multicolumn{1}{c}{03.11,~96.12,~00.77} \\\cline{2-14} 
 & \multicolumn{1}{c}{64} & \arrvline & \multicolumn{1}{c}{12.16} & \multicolumn{1}{c}{239.23} & 
 \multicolumn{1}{c}{10.80} &
 \multicolumn{1}{c}{8.22} &
 \multicolumn{1}{c}{0,~1,~16} & \arrvline & \multicolumn{1}{c}{10.95} & \multicolumn{1}{c}{14.44} & \multicolumn{1}{c}{7244} & \multicolumn{1}{c}{5.59} & \multicolumn{1}{c}{06.75,~83.40,~09.86} \\\cline{2-14}
 & \multicolumn{1}{c}{512} & \arrvline & \multicolumn{1}{c}{12.16} & \multicolumn{1}{c}{239.19} & 
 \multicolumn{1}{c}{10.92} &
 \multicolumn{1}{c}{8.23} &
 \multicolumn{1}{c}{0,~0,~17} & \arrvline & \multicolumn{1}{c}{10.90} & \multicolumn{1}{c}{5.04} & \multicolumn{1}{c}{22197} & \multicolumn{1}{c}{5.65} & \multicolumn{1}{c}{00.53,~99.16,~00.31} \\\hline\hline

 \multirow{3}{*}{S201} & \multicolumn{1}{c}{8} & \arrvline & \multicolumn{1}{c}{12.07} & \multicolumn{1}{c}{223.39} &
 \multicolumn{1}{c}{10.61} &
 \multicolumn{1}{c}{8.14} &
 \multicolumn{1}{c}{0,~0,~08} & \arrvline & \multicolumn{1}{c}{10.84} & \multicolumn{1}{c}{2.41} & \multicolumn{1}{c}{502} & \multicolumn{1}{c}{5.68} & \multicolumn{1}{c}{00.11,~99.86,~00.03} \\\cline{2-14} 
 & \multicolumn{1}{c}{64} & \arrvline & \multicolumn{1}{c}{12.07} & \multicolumn{1}{c}{222.11} & 
 \multicolumn{1}{c}{10.72} &
 \multicolumn{1}{c}{8.08} &
 \multicolumn{1}{c}{0,~0,~08} & \arrvline & \multicolumn{1}{c}{10.75} & \multicolumn{1}{c}{13.46} & \multicolumn{1}{c}{4458} & \multicolumn{1}{c}{5.73} & \multicolumn{1}{c}{03.97,~55.03,~41.00} \\\cline{2-14}
 & \multicolumn{1}{c}{512} & \arrvline & \multicolumn{1}{c}{12.04} & \multicolumn{1}{c}{218.37} & 
 \multicolumn{1}{c}{10.65} &
 \multicolumn{1}{c}{8.09} &
 \multicolumn{1}{c}{0,~0,~09} & \arrvline & \multicolumn{1}{c}{10.49} & \multicolumn{1}{c}{18.47} & \multicolumn{1}{c}{14437} & \multicolumn{1}{c}{5.48} & \multicolumn{1}{c}{00.53,~99.16,~00.31} \\\hline\hline

 \multirow{3}{*}{S221} & \multicolumn{1}{c}{8} & \arrvline & \multicolumn{1}{c}{11.97} & \multicolumn{1}{c}{206.19} &
 \multicolumn{1}{c}{10.58} &
 \multicolumn{1}{c}{7.79} &
 \multicolumn{1}{c}{0,~0,~12} & \arrvline & \multicolumn{1}{c}{10.72} & \multicolumn{1}{c}{18.52} & \multicolumn{1}{c}{300} & \multicolumn{1}{c}{5.38} & \multicolumn{1}{c}{00.72,~99.20,~00.07} \\\cline{2-14} 
 & \multicolumn{1}{c}{64} & \arrvline & \multicolumn{1}{c}{11.98} & \multicolumn{1}{c}{208.47} & 
 \multicolumn{1}{c}{10.62} &
 \multicolumn{1}{c}{8.12} &
 \multicolumn{1}{c}{0,~0,~08} & \arrvline & \multicolumn{1}{c}{10.77} & \multicolumn{1}{c}{9.70} & \multicolumn{1}{c}{3268} & \multicolumn{1}{c}{5.51} & \multicolumn{1}{c}{00.93,~98.92,~00.15} \\\cline{2-14}
 & \multicolumn{1}{c}{512} & \arrvline & \multicolumn{1}{c}{11.98} & \multicolumn{1}{c}{207.87} & 
 \multicolumn{1}{c}{10.70} &
 \multicolumn{1}{c}{7.83} &
 \multicolumn{1}{c}{0,~0,~10} & \arrvline & \multicolumn{1}{c}{10.70} & \multicolumn{1}{c}{18.27} & \multicolumn{1}{c}{29047} & \multicolumn{1}{c}{5.45} & \multicolumn{1}{c}{04.78,~89.42,~05.80} \\\hline\hline

 \multirow{3}{*}{S264} & \multicolumn{1}{c}{8} & \arrvline & \multicolumn{1}{c}{11.95} & \multicolumn{1}{c}{203.88} &
 \multicolumn{1}{c}{10.67} &
 \multicolumn{1}{c}{7.96} &
 \multicolumn{1}{c}{0,~0,~11} & \arrvline & \multicolumn{1}{c}{10.59} & \multicolumn{1}{c}{10.74} & \multicolumn{1}{c}{405} & \multicolumn{1}{c}{5.49} & \multicolumn{1}{c}{02.09,~97.89,~00.02} \\\cline{2-14} 
 & \multicolumn{1}{c}{64} & \arrvline & \multicolumn{1}{c}{11.95} & \multicolumn{1}{c}{203.36} & 
 \multicolumn{1}{c}{10.69} &
 \multicolumn{1}{c}{8.00} &
 \multicolumn{1}{c}{0,~0,~09} & \arrvline & \multicolumn{1}{c}{10.44} & \multicolumn{1}{c}{46.97} & \multicolumn{1}{c}{1239} & \multicolumn{1}{c}{4.95} & \multicolumn{1}{c}{10.00,~78.34,~11.66} \\\cline{2-14}
 & \multicolumn{1}{c}{512} & \arrvline & \multicolumn{1}{c}{11.94} & \multicolumn{1}{c}{200.93} & 
 \multicolumn{1}{c}{10.74} &
 \multicolumn{1}{c}{8.10} &
 \multicolumn{1}{c}{0,~0,~10} & \arrvline & \multicolumn{1}{c}{10.25} & \multicolumn{1}{c}{36.68} & \multicolumn{1}{c}{8515} & \multicolumn{1}{c}{5.14} & \multicolumn{1}{c}{12.58,~66.40,~21.02} \\\hline
\end{tabular}
\caption{A summary of selected global properties of our sample at $z=0$. The first two columns show the halo number and refinement factor (RF). The next five columns show quantities not directly related to CGM gas: halo mass, virial mass, galaxy stellar mass, central supermassive black hole mass, and number of mergers since $z=2$, split based on merger mass ratio. The last five columns show quantities related to CGM gas: total CGM gas mass, cold gas mass fraction, number of cold CGM gas clouds, average CGM gas temperature, and volume fractions of the three phases of CGM gas. Refer to main text for details.}
\label{table:halo_summary}
\end{table*}

\begin{figure*}
\centering 
\includegraphics[width=8cm]{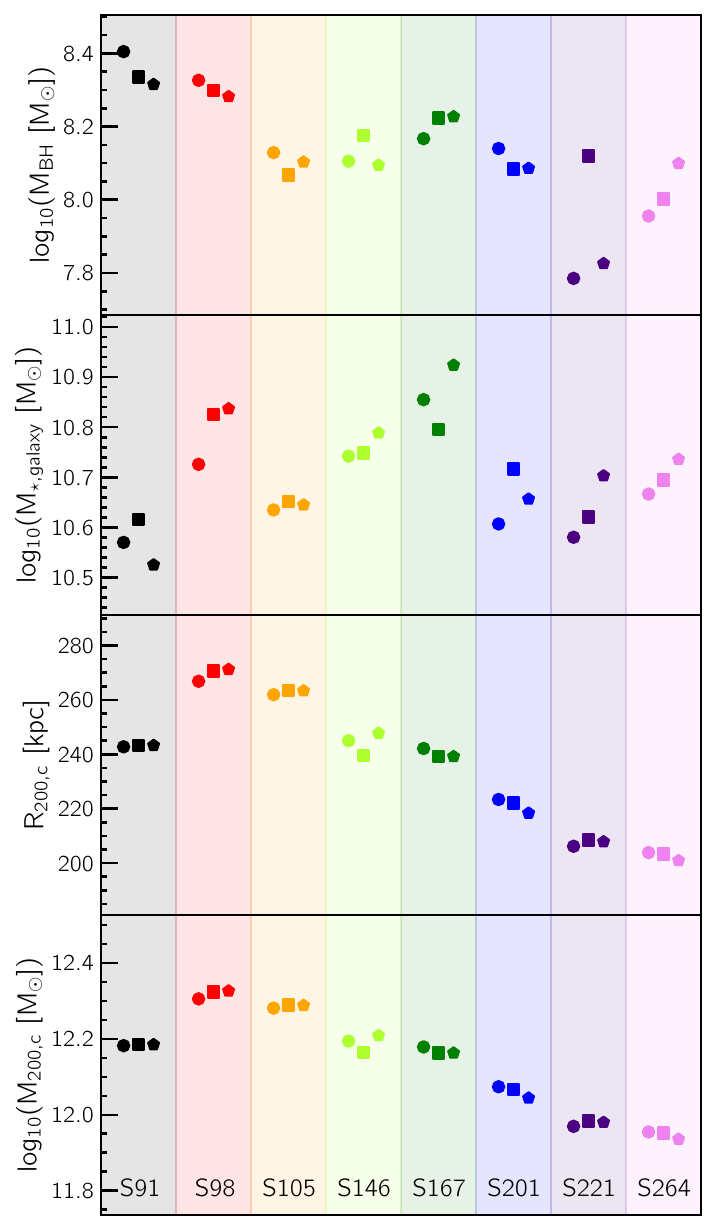}
\includegraphics[width=8cm]{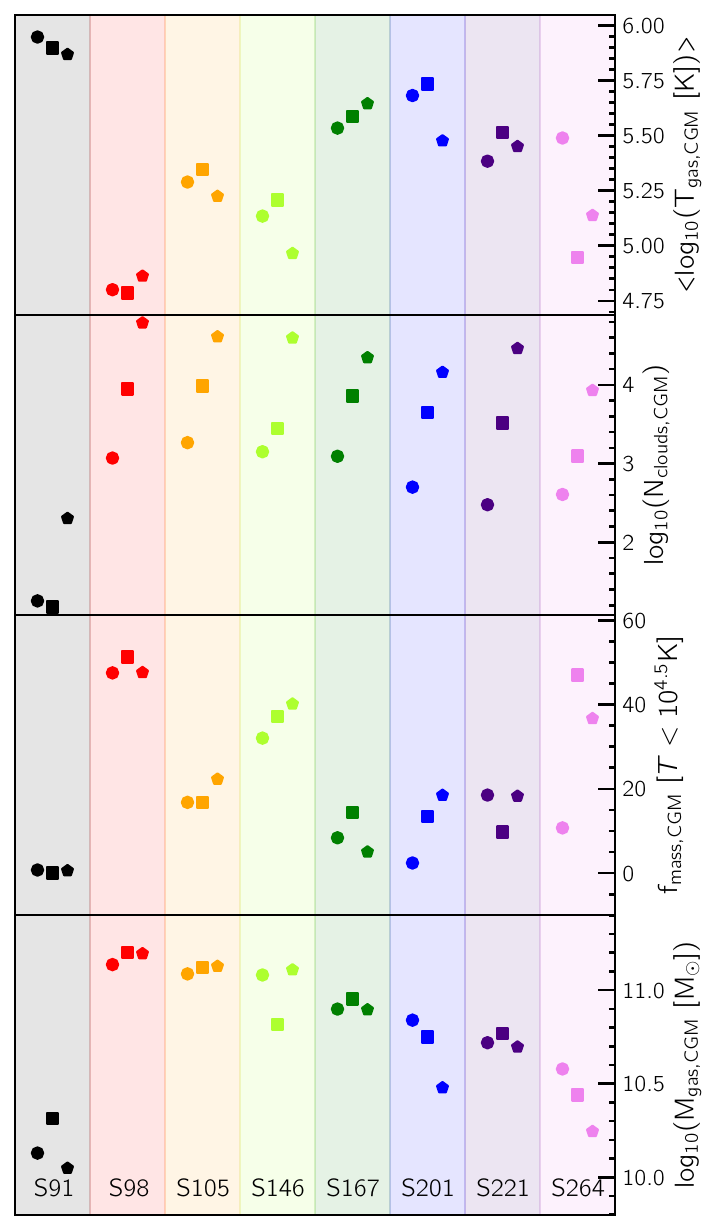}
\caption{Resolution trends for physical properties of our simulated galaxies and their gaseous halos, taken as a subset from Table~\ref{table:halo_summary}. Each color corresponds to a given halo. Runs at RF8, RF64 and RF512 are represented using circles, squares and pentagons, respectively. While all quantities fluctuate with changing resolution, due to inherent stochasticity and timing offset effects, no global resolution trend is present, except for the case of CGM gas clouds.}
\label{fig:table_sum}
\end{figure*}


\subsection{A comparison with other projects}\label{differences_analogs}

Before proceeding further, we compare our simulation(s) versus several other similar projects, including analogous CGM refinement techniques. Figure~\ref{fig:comp_projects} shows the (median) spatial resolution of simulated halos as a function of galactocentric distance normalised by the virial radius. The main panel only includes cold ($T < 10^{4.5}$~K) gas, while the inset shows results for all gas, with satellite gas excised in both cases. Note that the inset shows only a subset of curves from the main panel.

The three green curves correspond to $z=0$ Milky Way-like halos selected from the three TNG boxes. These are all large-volume uniform-resolution boxes, run at different resolutions: TNG50-1 (dashed; $m_b$\,$\sim$\,$8 \times 10^4$~M$_\odot$), TNG100-1 (dot-dashed; $m_b$\,$\sim$\,$10^6$~M$_\odot$) and TNG300-1 (dotted; $m_b$\,$\sim$\,$10^7$~M$_\odot$). The blue curves show examples of projects that aim to substantially improve numerical resolution: the dashed blue curve corresponds to Auriga Level-2 \citep{grand2021}, the (currently) best-resolved zoom-in simulation of a Milky Way-like galaxy, in which the mass resolution of gas is constant throughout the halo ($m_b$\,$\sim$\,$800$~$\rm{M}_\odot$). The dot-dashed curve shows CGOLS \citep{schneider2018}, an idealised $10 \times 10 \times 20$~kpc$^3$ simulation of a $M_\star$\,$\sim$\,$10^{10}$\,$\rm{M_\odot}$ disk galaxy simulated at a constant $\sim$\,$5$~pc spatial resolution throughout.

As opposed to the above simulations that maintain a fixed resolution, either mass- or spatial-, throughout the region of interest, the black curves show examples of CGM-refinement simulations, wherein resolution is preferentially increased within a refinement region that traces the CGM. The solid line shows one of the GIBLE-RF512 halos. The ($z$\,$\sim$\,$2$) $10^{12}~\rm{M_\odot}$ halo from \cite{suresh2019}, where a similar refinement technique was used as GIBLE to attain a CGM gas mass resolution of $\sim$\,$2200$\,M$_\odot$, is shown through the dashed curve. The dot-dashed curve shows TEMPEST \citep{hummels2019}, a simulation of a Milky Way-like galaxy at $z=0$ run with the grid code \textsc{enzo} \citep{bryan2014}, with enhanced spatial resolution of $\sim$\,$500$~pc within a $200$~kpc box centred on the galaxy, and coarser resolution at farther distances. The FOGGIE simulations \citep{peeples2019} use a similar technique as TEMPEST, and are not shown explicitly. The halo from \cite{voort2019} is shown through the dotted curve. This is a zoom-in simulation of a Milky Way-like galaxy run with \textsc{arepo} at a base resolution of $m_b$\,$\sim$\,$5 \times 10^4 \rm{M_\odot}$, i.e. similar to TNG50-1, but with added spatial refinement that sets a maximum spatial resolution for gas of 1kpc. 

As expected, the zoom-in simulations achieve much better resolutions in comparison to the large volume boxes, although at the expense of a smaller sample size. With the CGM-refinement technique, it is possible to simulate a relatively large sample of galaxies at super-high (CGM gas) resolution. For instance, the current sample of GIBLE is composed of eight halos that are comparable in (CGM) resolution to Auriga Level-2, of which there is only one galaxy. The obvious caveat is that these simulations are intended primarily to analyse the CGM. Moreover, as we discuss below, the choice of the CGM refinement technique is important, and needs to be made based on the primary aim of the project.

Improved mass resolution within the CGM, like that performed here in GIBLE and \cite{suresh2019}, has a greater impact on the cold phase, since this is the most dense. For instance, as seen in the main panel, the GIBLE-RF512 simulations resolve cold gas to a better extent than TEMPEST throughout the halo. However, when other phases of CGM-gas are included (inset), TEMPEST out performs GIBLE at almost all distances larger than $\sim$\,$0.3$\,$R_{\rm{200,c}}$.

Improved CGM-spatial refinement thus clearly effects the hot, volume-filling diffuse phase of gas to a larger degree. This is also well seen when the \cite{voort2019} curves are contrasted against TNG50, both of which have similar mass resolutions, but the former includes added spatial refinement. The TNG50 and \cite{voort2019} curves are comparable out to large distances in the main panel, since the size of cold gas cells at these distances is already smaller than or comparable to the 1kpc enforced maximum. However, the difference between the two is much more pronounced once the warm- and hot-phases of gas are included (inset), wherein all gas beyond $\sim$\,$0.3$\,$R_{\rm{200,c}}$ is significantly better resolved.

Our method is hence more advantageous for studying the clumpy, dense, cold phase of gas, which is believed to form fragmented structures down to the scale of a few parsecs \citep[e.g.][]{sparre2019}. Moreover, for the cold phase in general, a greater number of resolution elements would help improve the accuracy of the interaction with the ambient medium \citep[e.g.][]{goldsmith2016}. In order to address the puzzle of the small-scale cold phase CGM in a cosmological context, we therefore prefer the approach used here in GIBLE. However, for projects aimed at studying finer structures in the warm-hot phase, e.g. to study small scale turbulence of hot halo gas \citep[e.g.][]{schmidt2021}, or for better resolving the interaction of high-temperature, low-density outflows with their surroundings \citep[e.g.][]{smith2023}, adopting an improved spatial resolution throughout the region of interest would be more appropriate.

\section{Results}\label{results}

\subsection{An Overview of Global Properties}

We begin with a tabulation of several important global properties of our sample (at $z=0$) in Table~\ref{table:halo_summary}. The first column shows the different halo IDs (in TNG50-2) from which initial conditions were derived. Each halo is split into three sub-rows, corresponding to the different Refinement Factor (RF) choices, shown in the second column. The rest of the table is split into two parts. The left half shows quantities not directly associated with CGM gas: the first four columns give key properties used to describe galaxies and halos: from left to right, these are the halo virial mass ($M_{\rm{200,c}}$), halo virial radius ($R_{\rm{200,c}}$), galaxy stellar mass\footnote{Computed through the sum of masses of stars within an aperture of 30kpc with respect to the centre of the galaxy.} ($M_\star$) and the mass of the central supermassive black hole ($M_{\rm{BH}}$). Note that the three masses are shown in log$_{10}$ units. The next column briefly summarises the merger history of these galaxies, with the three values corresponding to the number of major, minor and all mergers since $z=2$. Major mergers are defined as those with merger mass ratios $\mu > 0.25$, while minor mergers are those with $0.1 < \mu < 0.25$ \citep[for more details, see][]{rodriguez2016}.

As galaxies in our sample are selected based on their similarity to the real Milky Way, halo- and galaxy stellar-masses are confined to relatively narrow bins, i.e. by construction (see Section~\ref{methods}). As a result, black hole masses and the merger histories of these galaxies are not too dissimilar as well (see \citealt{pillepich2023} for details on the level of variance expected within such a constrained sample).

\begin{figure*}
\centering 
\includegraphics[width=4cm]{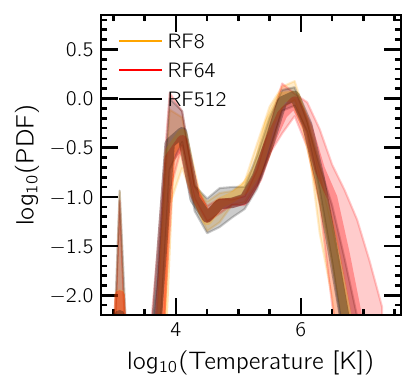}
\includegraphics[width=3.9cm]{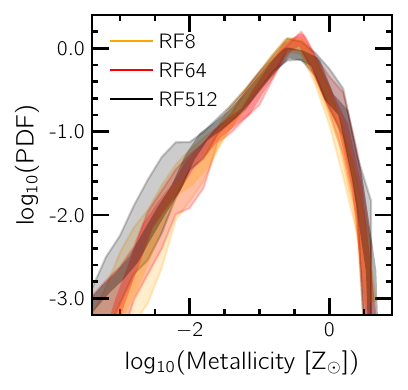}
\includegraphics[width=4cm]{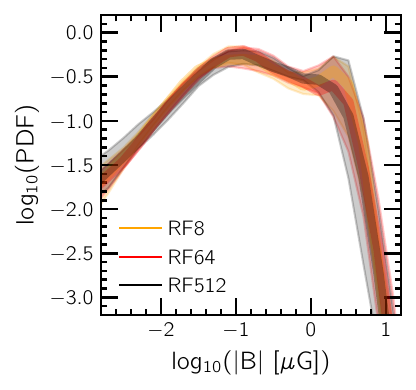}
\includegraphics[width=4cm]{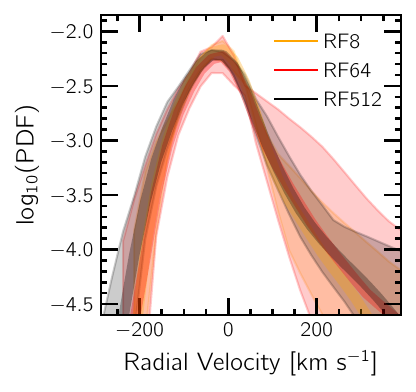}
\caption{Probability distribution functions (PDFs) of four different properties of CGM gas. For each halo in our sample, we compute (time) averages over ten snapshots ($\sim$~$300$~Myr) starting at $z=0$. The solid curves show the median of these time-averaged curves, while the shaded regions show the $16^{\rm{th}}$-$84^{\rm{th}}$ percentile variation of the sample. Results of the RF8, RF64 and RF512 runs are shown in orange, red and black, respectively. Although these time- and sample-averaged PDFs are converged with resolution, this does not necessarily imply that all results are also converged, as explored in detail throughout the rest of the paper.}
\label{fig:nonVaryingHist}
\end{figure*}

The next part of the table shows values related to CGM gas: the total CGM gas mass ($\rm{M_{g, CGM}}$), mass fraction of cold gas in the CGM (f$_{\rm{m, CGM, cold}}$), number of discrete, cold CGM gas clouds (N$_{\rm{cld, CGM}}$), average temperature of CGM gas ($\rm{\overline{T_{CGM}}}$)\footnote{The temperature of star-forming gas in the TNG model represents the effective state of the two phases. To overcome this limitation, we always set the temperature of star forming gas to $10^3$K, its cold phase value.}, and volume fractions of CGM gas split into three different phases (f$_{\rm{vol, CGM}}$), respectively, from left to right. Gas is split into three phases based on temperatures cutoffs: cold ($T<10^{4.5}$~K), warm ($10^{4.5}<T<10^{5.5}$~K) and hot ($T>10^{5.5}$~K). Following \cite{ramesh2023b}, we define cold gas clouds as contiguous sets of cold gas cells \citep[see also][]{nelson2020}. We consider all clouds down to the resolution limit, i.e. the minimum number of cells per cloud is set to one. 

Despite a relatively narrow bin in halo masses, properties of CGM gas show a large variation across the sample of eight galaxies \citep[see also][]{ramesh2023a}. For instance, the coldest halo in the sample has an average temperature of $\sim$\,$10^{4.8}$~K, more than an order of magnitude cooler than the hottest ($\sim$\,$10^{5.9}$~K). This results in a large variation of cold gas mass fractions, from as low as $\lesssim 1\%$ to as high as $\sim 50 \%$. 

To visually demonstrate the effect of numerical resolution, we plot a subset of these quantities in Figure~\ref{fig:table_sum}. The left half of the figure shows the first four columns of the first part of Table~\ref{table:halo_summary}, while the first four columns of the second part of the table are shown on the right panel. Each row corresponds to a single physical quantity. Both panels are split into eight shaded regions, with each color corresponding to a given halo. Runs at RF8, RF64 and RF512 resolutions are shown through circles, squares and pentagons respectively. 

Overall, halo mass shows the least variation with resolution, differing by at most $\sim$\,$0.03$~dex between runs of different resolutions. Galaxy stellar masses and BH masses show larger variations ($\lesssim \pm$\,$0.1$~dex), as do some of the gas quantities, like CGM gas mass ($\lesssim \pm$\,$0.2$~dex), cold gas mass fractions ($\lesssim 20-30$~$\%$) and average halo temperatures ($\lesssim \pm 0.3$~dex). However, no global monotonic trend is apparent for any of these quantities. This suggests that such variations are likely a result of the randomness of the butterfly effect \citep{genel2019} and/or the zoom-timing effect \citep{springel2008}, and not directly sensitive to changing numerical resolution.

The one quantity for which a clear resolution trend is visible is the number of discrete, cold CGM gas clouds. For all but the first halo (S91), a greater number of clouds are found at higher resolution. As discussed in \cite{ramesh2023b}, this is a direct result of the existence of a resolution limit for every simulation. Structures smaller than the resolution limit are simply absent. We return to this discussion in Figure~\ref{fig:cloudMass}.

In Figure~\ref{fig:nonVaryingHist}, we further assess numerical convergence of overall-halo gas properties by studying probability distribution functions (PDFs). For each halo, we compute averages over ten snapshots ($\sim$~$300$~Myr) starting at $z=0$. Time-averaging over such a relatively-short interval helps suppress the impact of the butterfly effect and the timing problem discussed above, while also ensuring that the dynamical evolution of the halo does not play a major role. In each panel, the solid curves show the median of each time-averaged halo-curve, while the shaded regions shown the variation across the sample. Results from RF8, RF64 and RF512 are shown in orange, red and black, respectively.

From left to right, panels show the PDFs of gas temperature, metallicity, magnetic field strength and radial velocity, respectively. In each case, the medians appear to be converged with resolution, at least at the level of resolution considered here. However, as we will explore throughout the remainder of the paper, this does not imply that all quantities related to CGM gas are necessarily converged, especially those of small-scale properties, as has already been seen with the increase in number of cold clouds with improving resolution, despite the PDFs of gas temperature being well converged.

\subsection{HI Covering Fractions of Halo Gas}

The covering fraction of a given species is a commonly used observational metric to quantify the distribution of halo gas when viewed in 2D projection. Here, we make predictions for the covering fractions (at $z=0$) of commonly studied ions, and explore their sensitivity to numerical resolution. 

As we did not save the abundances of specific metal species in the GIBLE simulation outputs, we assume that hydrogen contributes $76\%$ to the mass of each gas cell, and adopt solar abundance for all other metal species. When estimating the fraction of HI that comprises the (neutral) hydrogen-mass of the cell, we use the model of \cite{gnedin2011} to excise the contribution of H$_2$ \citep[following][]{popping2019}. We use \textsc{cloudy} \citep[][v13.03]{ferland2013} to model the abundances of various ionization states of the different metal species \citep[following][]{nelson2018b,ramesh2023a}.

\begin{figure*}
\centering 
\hspace{0.94cm}
\includegraphics[width=15.97cm]{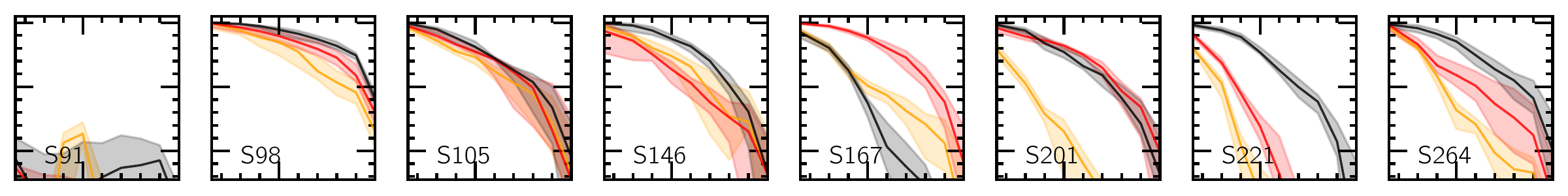}
\includegraphics[width=17.55cm]{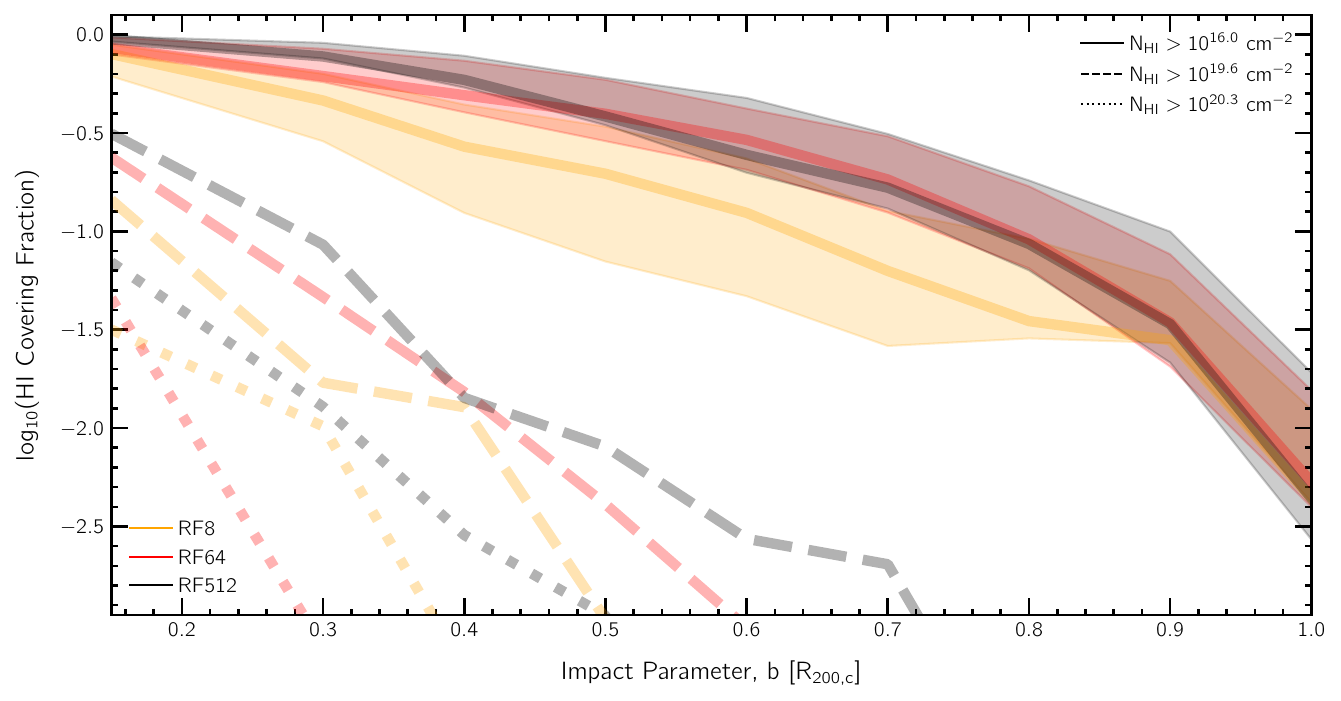}
\caption{Trends of HI covering fractions as a function of impact parameter. In the main panel, we show median curves for the sample, split based on numerical resolution (different line colors) and N$_{\rm{HI}}$ cutoffs (different line types). Shaded regions, shown only for the $10^{16}~\rm{cm^{-2}}$ case, represents the halo-to-halo variation across the sample. Median covering fractions decrease with distance, and are lower for higher column density thresholds. Increased numerical resolution generally leads to higher covering fractions, except at the high column density end, where the impact of resolution is superseded by other stochastic effects. The small panels in the top row show the radial profiles of HI covering fraction for the eight individual halos in our sample. Refer to main text for details.}
\label{fig:coveringFracHI}
\end{figure*}

In Figure~\ref{fig:coveringFracHI}, we consider the covering fractions of HI (y-axis) as a function of (2D) impact parameter (normalised by $\rm{R_{200,c}}$; x-axis), a ubiquitous tracer of cold gas. The main panel considers three commonly used column density thresholds for HI: $>$\,$10^{16}~\rm{cm^{-2}}$ (solid; pLLSs), $>$\,$10^{19.6}~\rm{cm^{-2}}$ (dashed; sub-DLAs) and $>$\,$10^{20.3}~\rm{cm^{-2}}$ (dotted; DLAs). Results from the RF8, RF64 and RF512 runs are shown in orange, red and black, respectively. Each galaxy is considered in a random orientation, and curves show median trends of the sample. For the $>$\,$10^{16}~\rm{cm^{-2}}$ case alone, we show the halo-to-halo variation through the $16^{\rm{th}}$-$84^{\rm{th}}$ percentile shaded regions.

For a N$_{\rm{HI}}$ threshold of $10^{16}~\rm{cm^{-2}}$, median covering fractions are close to unity at the inner boundary of the CGM ($0.15~R\rm{_{200,c}}$), for all three resolution runs. For the best resolved run (RF512; black), there is a steady decrease to $\sim$\,$0.4$ at $0.5~R\rm{_{200,c}}$, followed by a steeper drop to $\sim$\,$0.005$ at the virial radius. At this N$_{\rm{HI}}$ threshold, results from the RF64 run (red) are more or less consistent with the black curve, except in the innermost region of the halo. For the RF8 run, however, covering fractions are systematically lower by $\sim$\,$0.1-0.3$~dex with respect to the black curve out to an impact parameter of $\sim$\,$0.9~R\rm{_{200,c}}$. Beyond this distance, in the outermost regions of the halo, all three curves are converged. The overlap of the black and red curves demonstrates that the HI covering fraction, at this column density threshold, is numerically converged already at our RF64 resolution. 

As one transitions towards higher N$_{\rm{HI}}$ thresholds, the dependence on resolution becomes stronger. At a cutoff of $10^{19.6}~\rm{cm^{-2}}$ (dashed lines), the median covering fraction of the RF512 run is $\sim$\,$0.3$ at $0.15~R\rm{_{200,c}}$, and is over two orders of magnitude smaller ($\sim$\,$0.002$) already at $0.7~R\rm{_{200,c}}$. Columns of gas at such density thresholds are thus rare to find at large impact parameters. For the RF64 run, the median is lower by $\sim$\,$0.1$~dex in the inner halo ($\lesssim$\,$0.3~R\rm{_{200,c}}$) with respect to RF512, and the offset grows beyond $\sim$\,$0.5~R\rm{_{200,c}}$. The RF8 run has even lower convering fractions, demonstrating that dense gas (at this N$_{\rm{HI}}$ threshold) is effected by numerical resolution, and results are not well converged, at least up to a resolution of $\sim$\,$10^4~\rm{M_\odot}$. 

For N$_{\rm{HI}}$\,$\gtrsim$\,$10^{20.3}~\rm{cm^{-2}}$ (dotted curves), median covering fractions drop further, for all resolution levels. For instance, in the RF512 runs, the median covering fraction is lower by $\sim$\,$0.6$~dex with respect to the black dashed curve. The red and orange curves are again vertically offset with respect to the black curve, although no monotonic trend with resolution is present here. This suggests that, at column density thresholds as high as $10^{20.3}~\rm{cm^{-2}}$, the impact of resolution is superseded by other effects, possibly by the choice of a random orientation for each galaxy, or due to the randomness of stochastic processes, as discussed above.

In the top panels, we show a collection of eight stamps, each corresponding to a given halo. Each stamp contains three curves, corresponding to the three different resolution runs, with the same color scheme as the main panel. For visual clarity, we show only the case with a N$_{\rm{HI}}$ threshold of $10^{16}~\rm{cm^{-2}}$. For each galaxy, we pick 100 random orientations, and show the median with solid lines. Shaded regions represent the $16^{\rm{th}}$-$84^{\rm{th}}$ percentile variation of the different orientations. As seen, at least at this column density threshold, the choice of orientation does not play a dominant role, with percentile regions being generally narrow ($\lesssim$\,$0.1$~dex).

While most individual halos are consistent with the findings of the main panel, there are exceptions. For instance, in the fourth case from the left, the RF64 curve lies below the RF8, while the fifth halo gas the RF512 curve below the other two. This variability demonstrates that it is impossible to accurately assess the impact of resolution based on a single halo, and a sample of multiple halos is needed to overcome uncertainties related to effects including numerical stochasticity and/or timing offsets.

Adding further complication, the HI content of gas in simulations is sensitive to the (possibly self-shielded) photo-ionization rate. Using a semi-empirical model, \cite{faerman2023} make predictions for the radial profile of HI column densities. In addition, they study the effect of varying the amount of non-thermal pressure support of cold gas ($\eta$). Greater values of $\eta$ correspond to higher rates of photo-ionisation. Out to impact parameters of $\sim$\,$0.5~R_{\rm{200,c}}$, they find that their results are sensitive to the chosen value of $\eta$. For instance, at an impact parameter of $0.2~R_{\rm{200,c}}$, their low- (high-)$\eta$ case predicts a median HI column density of $10^{17.5}$ ($10^{16.5}$)~$\rm{cm^{-2}}$, as a result of higher photo-ionization rates leading to a removal of HI. The choice of the (sub-grid) self-shielding model employed for dense gas in cosmological simulations is thus important, and predictions of quantities related to HI could possibly be sensitive to this choice. However, we have explicitly checked that trends of HI covering fraction are qualitatively similar to those of H covering fractions (not shown), suggesting that the self-shielding recipe employed by TNG \citep{rahmati2013} does not bias results. 

To provide an idea of the typical HI-densities and trends of covering fractions recovered from both observations and simulations, we quote some numbers from current literature. Note that most of these are not apple-to-apple comparisons, i.e the galaxy sample selections, redshift ranges, and other details are not replicated, and are intended to be taken at face value only.
Observationally, for a set of 32 low-$z$ COS-Halos galaxies, \cite{prochaska2017} find that the median N$_{\rm{HI}}$ is $\gtrsim$\,$10^{17.5}$\,$\rm{cm^{-2}}$ at impact parameters $<$\,$60$~kpc, which is consistent with our RF64 and RF512 curves that predicts a covering fraction close to unity for pLLS systems at impact parameters $\lesssim$\,$0.3~R_{\rm{200,c}}$. For a sample of 32 strong Ly-$\alpha$ absorbers in the redshift range $0.2$\,$\lesssim$\,$z$\,$\lesssim$\,$1.4$, \cite{weng2023} too infer a trend of decreasing N$_{\rm{HI}}$ with increasing impact parameter, with values dropping from $\sim$\,$10^{19.5}$\,$\rm{cm^{-2}}$ at $0.15~R_{\rm{200,c}}$ to $\sim$\,$10^{18.}$\,$\rm{cm^{-2}}$ at the virial radius. Using a sample of 32 $z=2$ N$_{\rm{HI}} \simeq $\,$10^{17.2}$\,$\rm{cm^{-2}}$ absorbers, \cite{rubin2015} find that the covering fraction is $\sim$\,$0.6-0.8$ at impact parameters $\lesssim$\,$150$~kpc, and drops to $\sim$\,$0.2$ at larger distances. \cite{rudie2012} find that, for a sample of 886 $z$\,$\sim$\,$2-3$ galaxies, the covering fraction for N$_{\rm{HI}}$\,$>$\,$10^{16}$\,$\rm{cm^{-2}}$ absorbers varies between $0.6$ in the inner most part of the halo and $\lesssim$\,$0.1$ at larger impact parameters. These trends of decreasing covering fractions of HI with increasing impact parameter is qualitatively consistent with our GIBLE curves.

\begin{figure*}
\centering 
\includegraphics[width=5.7cm]{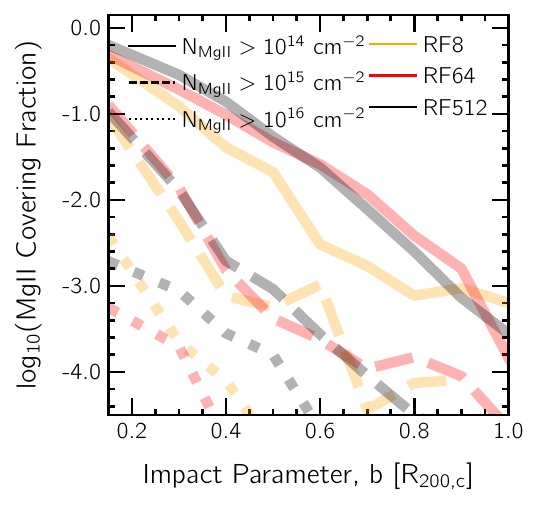}
\includegraphics[width=5.8cm]{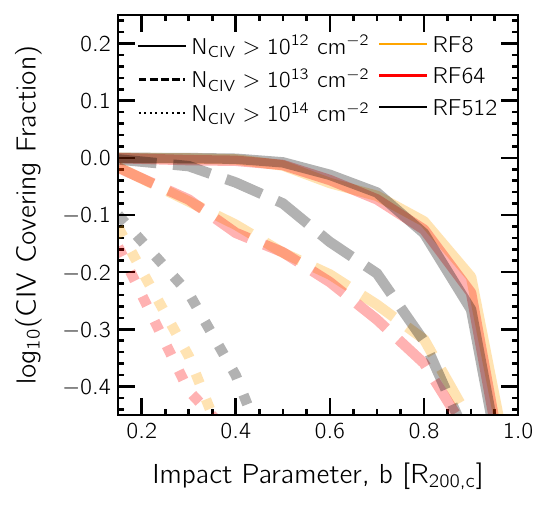}
\includegraphics[width=5.8cm]{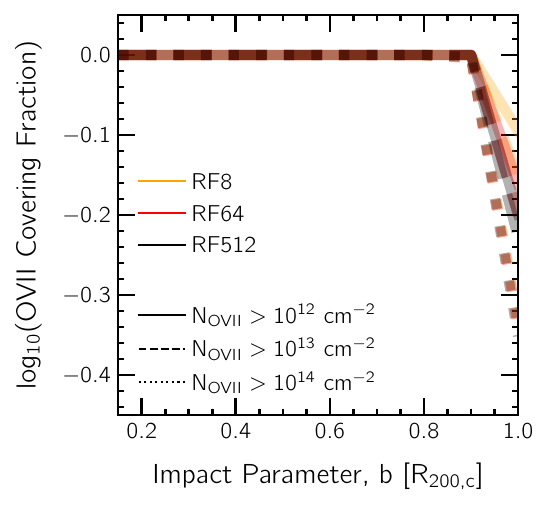}
\caption{Trends of covering fractions of MgII (left), CIV (centre) and OVII (right) as a function of impact parameter. In each panel, the colors represent the different resolution runs, while line styles correspond to different column density thresholds for each ion. Overall, the hot phase, traced by OVII, is the best converged (numerically) in our simulations, while the convergence of the other two ions depend on the column density thresholds employed.}
\label{fig:coveringFrac}
\end{figure*}

Using standard cosmological zoom simulations, \cite{hani2019} also study the distribution and covering fractions of HI and various ionic species. They adopt a sample of 40 $z=0$ MW-like galaxies from the Auriga project simulated with an average gas resolution of $\sim$\,$5 \times 10^4~\rm{M_\odot}$, i.e. at a resolution roughly between our RF8 and RF64 runs. For N$_{\rm{HI}}$\,$\gtrsim$\,$10^{14.15}~\rm{cm^{-2}}$ absorbers, they find that the distribution of covering fractions is large, for sightlines stacked between $0.2~R_{\rm{200,c}}$ and $R_{\rm{200,c}}$: while the distribution peaks at $\sim$\,$0.65$, values range between $\sim$\,$0$ and $\sim$\,$0.8$. They find that this large variation is a result of the $\sim$\,$1$~dex scatter in stellar mass, and a wide distribution of disk fractions and AGN luminosity across their sample.

The impact of resolution on HI covering fractions has also been explored by the other `CGM refinement' projects mentioned earlier. For a set of $z$\,$\sim$\,$2$ MW-like progenitors from the FOGGIE simulation project, \cite{peeples2019} find that their method leads to higher covering fractions for column densities $10^{15}$\,$\lesssim$\,N$_{\rm{HI}}$\,$\lesssim$\,$10^{17}~\rm{cm^{-2}}$. Similarly, \cite{voort2019} show that increased spatial refinement leads to larger HI covering fractions for $10^{14}$\,$\lesssim$\,N$_{\rm{HI}}$\,$\lesssim$\,$10^{19}~\rm{cm^{-2}}$, for a single MW-like analog at $z=0$, simulated at the same resolution as the Auriga sample discussed above, i.e. $m_b$\,$\sim$\,$5 \times 10^4~\rm{M_\odot}$ (but with added spatial refinement for a subset of runs). On the other hand, our results show decent convergence (for RF64 and above) in the low N$_{\rm{HI}}$ end ($\lesssim$\,$10^{19}~\rm{cm^{-2}}$). As discussed in Section~\ref{differences_analogs}, this difference is likely a consequence of the difference in our methods: increased spatial resolution largely impacts the volume filling hot phase, while our technique of increased mass resolution has a greater impact on the cold phase. Moreover, as previously discussed in \cite{suresh2019}, the `base' resolution plays an important role: in our RF64 and RF512 runs, the median spatial resolution of cold gas is already below 1~kpc, which is the point at which \cite{voort2019} find a difference with their added spatial resolution.

\subsection{Metal Covering Fractions}

In Figure~\ref{fig:coveringFrac}, we explore trends of covering fractions of commonly used metal ion tracers of multi-phase CGM gas \citep[see e.g.][]{donahue2022}: MgII (cold gas; left panel), CIV (warm gas; centre panel) and OVII (hot X-ray emitting gas; right panel). Each panel shows the covering fraction of the corresponding ion as a function of impact parameter (normalised by $\rm{R_{200,c}}$). The medians of the three different resolution runs are shown through different colors: orange (RF8), red (RF64) and black (RF512). The different line styles correspond to different column density thresholds, as elaborated below.

For a column density threshold of N$_{\rm{MgII}}$\,$\gtrsim$\,$10^{14}~\rm{cm^{-2}}$ (solid curves, left panel), the median covering fraction at $\sim$\,$0.15~R\rm{_{200,c}}$ is close to unity for the RF512 run, which reduces steadily to $<$\,$10^{-3}$ at the virial radius. Values of RF64 are consistent with RF512 at all distances, but the RF8 curve is vertically offset to lower covering fractions, almost at all distances. Covering fractions reduce with increasing thresholds on column density. At N$_{\rm{MgII}}$\,$\gtrsim$\,$10^{15}~\rm{cm^{-2}}$, (dashed) curves are vertically offset by $\sim$\,1~dex in comparison to the solid curves. Surprisingly, decent convergence is seen between resolution levels at this threshold, although curves are clearly noisy. For a higher threshold of N$_{\rm{MgII}}$\,$\gtrsim$\,$10^{16}~\rm{cm^{-2}}$, shown through dotted curves, covering fractions are over two orders of magnitudes smaller in comparison to the solid curves. No consistent trend with resolution is seen here, much like the HI case, which is expected since both these species largely trace the same gas.

These trends and results are largely in qualitative agreement with past observational and theoretical work.
For instance, using a large sample of $\sim$\,$160,000$ MgII absorbers in the redshift range $0.3$\,$\lesssim$\,$z$\,$\lesssim$\,$2.3$, \cite{anand2021} study the covering fractions in the CGM around luminous red galaxies (LRGs) and emission line galaxies (ELGs). At an equivalent width (EW) threshold of $0.4~A^\circ$, they find that the MgII covering fractions of LRGs (ELGs) is $\sim$\,$0.15$ ($0.6$) at an impact parameter of $20$~kpc, and drops to $\sim$\,$0.03$ ($0.04$) at $400$~kpc. At a higher EW threshold of $2.0~A^\circ$, these numbers reduce to $\sim$\,$0.05$ ($0.1$) at $20$~kpc, and $\sim$\,$0.001$ ($0.003$) at $400$~kpc. With a set of $16$ LRGs at intermediate redshift ($0.21$\,$\lesssim$\,$z$\,$\lesssim$\,$0.55$), \cite{zahedy2019} estimate the MgII covering fraction to be $0.4-0.7$ ($0.0-0.2$) for impact parameters $\lesssim$\,$100$~kpc ($100-160$~kpc), for a column density threshold of N$_{\rm{MgII}}$\,$\gtrsim$\,$10^{13}~\rm{cm^{-2}}$.

Using the EAGLE simulations, \cite{ho2020} studied the covering fractions around a set of $z$\,$\sim$\,$0.3$ galaxies. For a threshold of N$_{\rm{MgII}}$\,$\gtrsim$\,$10^{11.5}~\rm{cm^{-2}}$, they find that the MgII covering fraction at $10^{10.5}$\,$\lesssim$\,$M_\star/\rm{M_\odot}$\,$\lesssim$\,$10^{11.0}$ is $\sim$\,$0.55$ ($0.2$) at an impact parameter of $50$ ($150$)~pkpc. Moreover, they find that when galaxies are oriented edge-on, sightlines closer to the minor axis have larger MgII covering fractions, as a result of the directional nature of outflows \citep[e.g][]{peroux2020,truong2021,ramesh2023c}. In addition, they study the effect of the star-formation status of the galaxy. At this mass range, MgII covering fractions do not correlate with SFR, but in galaxies with smaller stellar masses, covering fractions are typically smaller when the central galaxy is quenched. 

\cite{nelson2020} study the MgII covering fractions for a sample of TNG50 halos in the range $10^{13.2}$\,$\lesssim$\,$M_{\rm{200,c}}/\rm{M_\odot}$\,$\lesssim$\,$10^{13.8}$, stacked at three snapshots ($z=[0.4, 0.5, 0.6]$). They compare TNG50-1 with -2, -3 and -4 (8, 64 and 512 times lower mass resolution, respectively), and find that the covering fraction monotonically decreases with coarsening resolution, for a N$_{\rm{MgII}}$ cutoff of $10^{16}~\rm{cm^{-2}}$. At this threshold, we find no monotonic trend with resolution with GIBLE. This suggests that effects of numerical resolution on MgII covering fractions are more important at the resolution of TNG50-1 (i.e RF8) and below, and hence less apparent apparent when comparing between the higher resolution RF64 and RF512 runs.

\begin{figure*}
\centering 
\includegraphics[width=16cm]{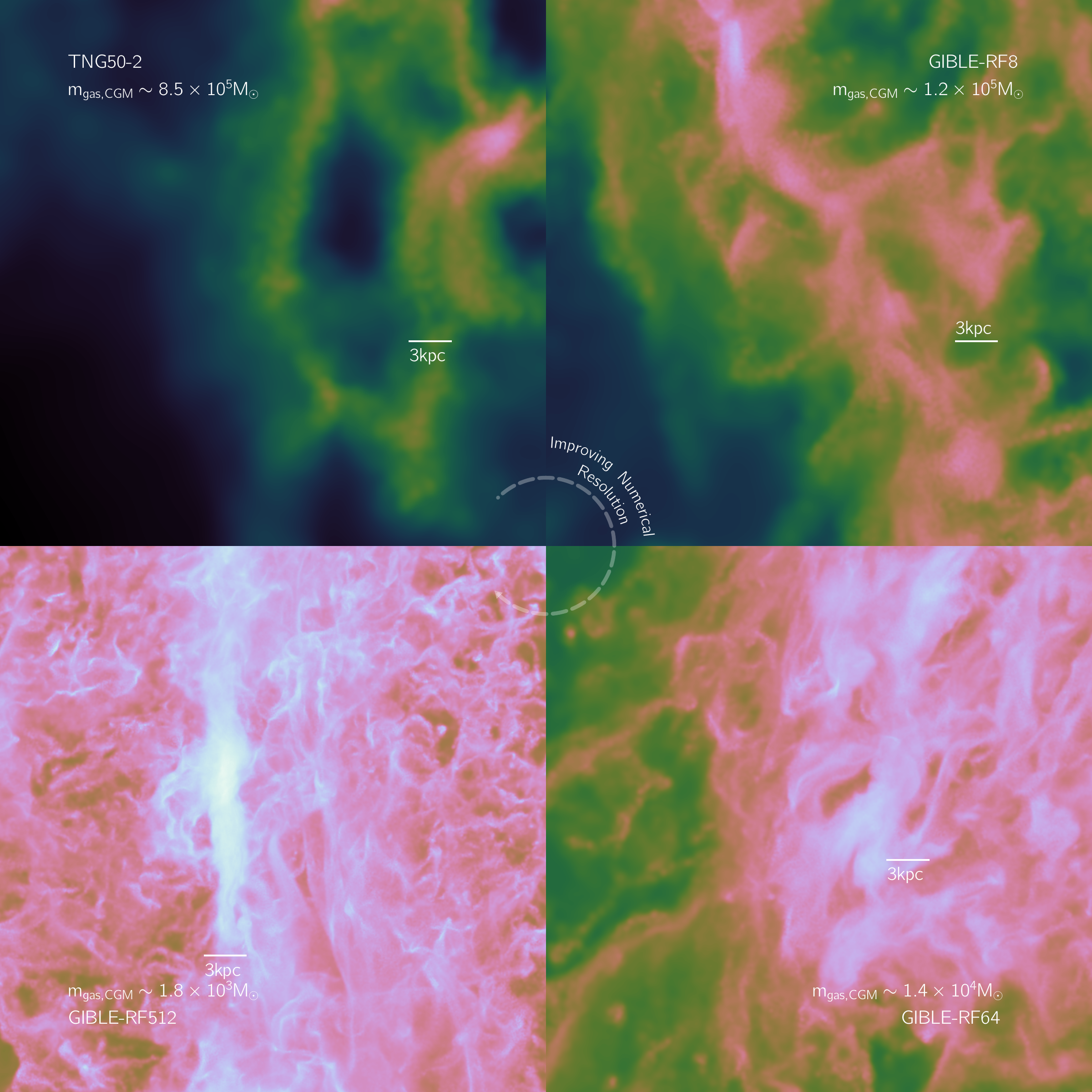}
\includegraphics[width=6cm]{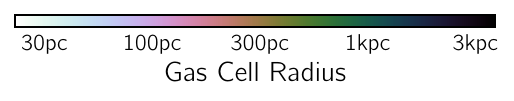}
\caption{A qualitative illustration of the impact of resolution on small scale structure. The layout is similar to that of Figure~\ref{fig:mainVisFigure}, wherein we show a zoomed-in region of one halo from four different simulations, with improving numerical resolution from top left to bottom left in clockwise direction. Colors correspond to the spatial resolution, i.e. radii of gas cells. An increased amount of small scale structure is visibly apparent at higher resolutions.}
\label{fig:mainVisFigureZoom}
\end{figure*}

\begin{figure*}
\centering 
\hspace{.95cm}
\includegraphics[width=16cm]{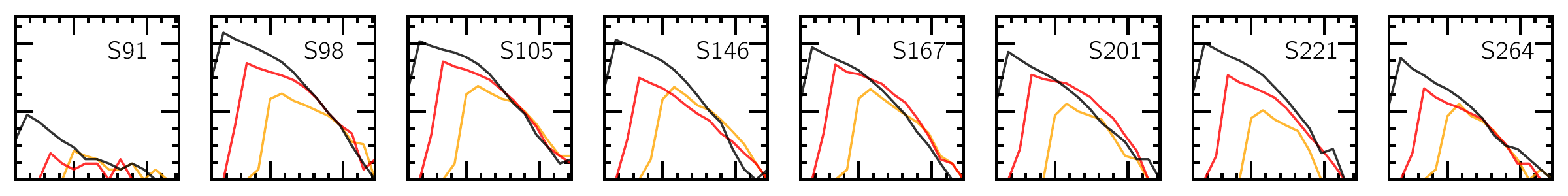}
\includegraphics[width=17.25cm]{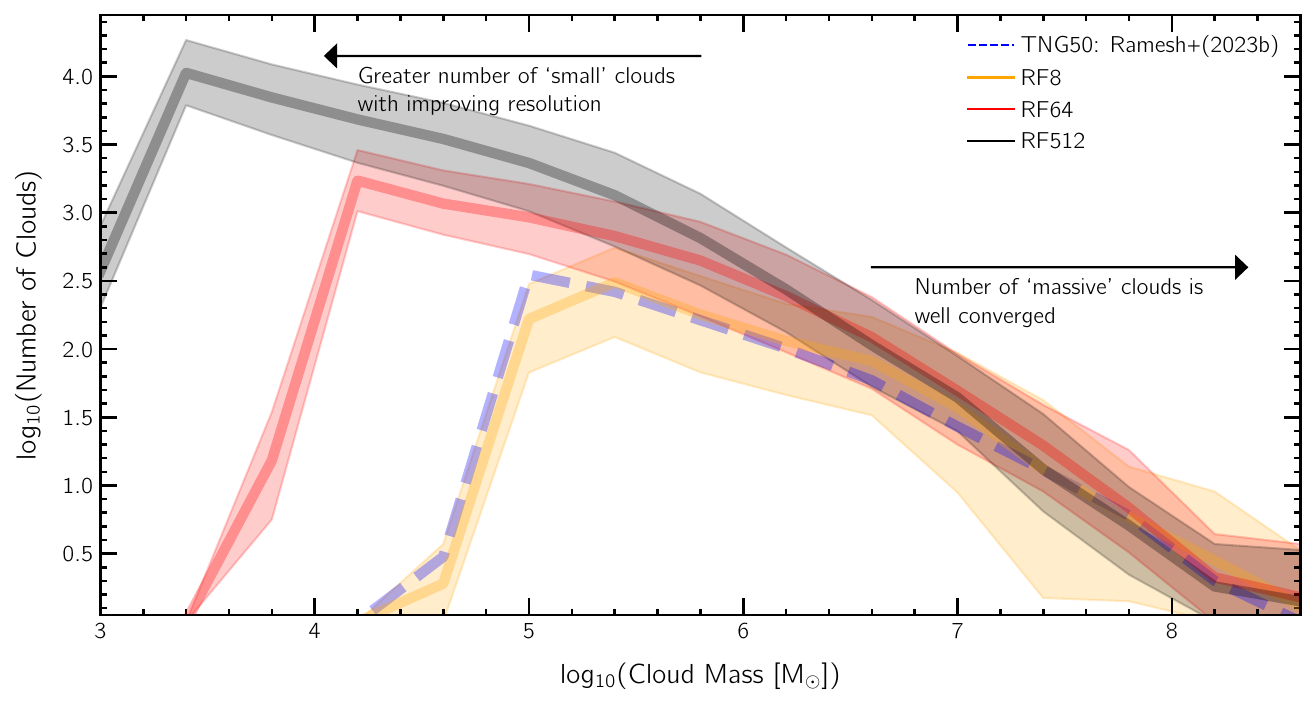}
\caption{Number of clouds as a function of their mass. The main panel shows the median of the sample with solid lines, while the shaded regions show the $16^{\rm{th}}$-$84^{\rm{th}}$ percentile values. The solid orange, red and black curves correspond to RF8, RF64 and RF512, respectively, while the dashed blue curve shows results from \protect\cite{ramesh2023b}, i.e. the entire TNG50 MW-like sample. Small panels in the top row show results for each of the eight GIBLE halos separately. Excellent convergence with numerical resolution is seen at the high mass end, while the number of small clouds increases with improving resolution, since these are simply absent in lower resolution runs.}
\label{fig:cloudMass}
\end{figure*}

Covering fractions of CIV, at a column density threshold of N$_{\rm{CIV}}$\,$\gtrsim$\,$10^{12}~\rm{cm^{-2}}$ (solid curves, centre panel) are relatively flat near unity between the inner boundary of the CGM ($\sim$\,$0.15~R\rm{_{200,c}}$) and $\sim$\,$0.5~R\rm{_{200,c}}$. At larger impact parameters, the value drops steadily, but is still as large as $\sim$\,$0.2$ at the virial radius. Numerical convergence, at this threshold, is good between the different runs. For N$_{\rm{CIV}}$\,$\gtrsim$\,$10^{13}~\rm{cm^{-2}}$ (dashed curves), the covering fractions for the RF512 run are offset by $\sim$\,$0.1$~dex with respect to the solid curve, except at impact parameters $\lesssim$\,$0.35~R_{\rm{200,c}}$. The RF8 and RF64 runs show slightly smaller values, possibly as a result of resolution effects kicking in. At even higher thresholds (N$_{\rm{CIV}}$\,$\gtrsim$\,$10^{14}~\rm{cm^{-2}}$; dotted curves), values are further smaller, and no monotonic trend is seen with numerical resolution. This marks the regime where stochastic effects overshadow any clear trend of resolution.

As before, we next provide a zeroth-order comparison with other results from the literature. 
Using 43 low-mass $z$\,$\lesssim$\,$0.1$ galaxies, \cite{boroldoi2014} showed that the CIV covering fraction for absorption systems with rest-frame equivalent width $W_r$\,$>$\,$0.1~A^\circ$ ($>$\,$0.3~A^\circ$) is $\sim$\,$0.8$ ($0.4$) at $0.2~R_{\rm{200,c}}$, and drops to $\sim$\,$0.4$ ($0.15$) at $0.4~R_{\rm{200,c}}$. \cite{schroetter2021} find that the covering fraction of CIV absorbers with $W_r$\,$>$\,$0.7~A^\circ$ at $z$\,$\sim$\,$1.2$ exceeds $50\%$ at impact parameters $\lesssim$\,$23^{+62}_{-16}$~kpc. With a sample of eight $L_\star$ galaxies observed at $z$\,$\sim$\,$2$, \cite{rudie2019} show that, for impact parameters $\lesssim$\,$100$~pkpc, eight, seven and six of those sightlines satisfy N$_{\rm{CIV}}$ thresholds of $10^{12}$, $10^{13}$ and $10^{14}~\rm{cm^{-2}}$, respectively. While direct comparison with our results is not straightforward, we note that these numbers are largely (qualitatively) consistent with the trends of our RF512 (black) curves. 

On the theoretical end, the semi-analytic model of \cite{faerman2023} predicts that, at fixed distance, the median CIV column density can vary between $\sim$\,$10^{12.5}~\rm{cm^{-2}}$ to $\sim$\,$10^{13.5}~\rm{cm^{-2}}$ out to impact parameters of $\sim$\,$0.5~R_{\rm{200,c}}$, depending on the chosen value of $\eta$ (see discussion above). As with HI, the variation is a result of greater photo-ionisation producing larger amounts of CIV. At larger impact parameters, their model predicts that N$_{\rm{CIV}}$ is roughly independent of $\eta$, and drops from  $\sim$\,$10^{12.5}~\rm{cm^{-2}}$ at $\sim$\,$0.5~R_{\rm{200,c}}$ to $\lesssim$\,$10^{12.0}~\rm{cm^{-2}}$ at the virial radius. These numbers are largely consistent with the trend of our solid curves (N$_{\rm{CIV}}$\,$>$\,$10^{12}~\rm{cm^{-2}}$).

With the same set of Auriga galaxies discussed above, \cite{hani2019} studied the covering fractions of N$_{\rm{CIV}}$\,$\gtrsim$\,$10^{14.7}~\rm{cm^{-2}}$ absorbers in the CGM. They find that the distribution of covering fractions is broad, varying from $\sim$\,$0$ to $\sim$\,$0.8$, with the distribution peaking at $\sim$\,$0.6$, when all sightlines between $0.2~R_{\rm{200,c}}$ and $R_{\rm{200,c}}$ are stacked. For these column densities, we find such high covering fractions only in the innermost region of the CGM ($\lesssim$\,$0.3~R_{\rm{200,c}}$).

Similarly, for a stacked set of sightlines passing through the CGM, the FOGGIE simulations predict that the covering fraction of CIV is $\sim$\,$40\%$ ($\sim$\,$5\%$) for N$_{\rm{CIV}}$\,$>$\,$10^{12}~\rm{cm^{-2}}$ ($>$\,$10^{14}~\rm{cm^{-2}}$) absorbers, albeit at $z=2$ \citep{peeples2019}. They find that increased resolution leads to higher covering fractions for $10^{11.5}$\,$\lesssim$\,N$_{\rm{CIV}}$\,$\lesssim$\,$10^{13.5}~\rm{cm^{-2}}$. However, our results are converged for lower thresholds (N$_{\rm{CIV}}$\,$\sim$\,$10^{12}~\rm{cm^{-2}}$), and the impact of increasing resolution is only seen at higher values (N$_{\rm{CIV}}$\,$\gtrsim$\,$10^{13}~\rm{cm^{-2}}$). As with HI, this is likely a direct consequence of differences in our methods as well as `base' resolution (see discussion above).

For the same column density cuts, we compare the covering fractions of OVII in the right panel, a tracer of the hotter, volume-filling component of the CGM. At all considered cutoffs, the covering fractions are close to unity, except at the very outskirts of the halo ($\gtrsim$\,$0.9~R_{\rm{200,c}}$). This is consistent with results from the SIMBA and EAGLE simulations, where covering fractions of OVII are close to $\sim$\,$100\%$ for thresholds N$_{\rm{OVII}}$\,$\lesssim$\,$10^{14.0}~\rm{cm^{-2}}$ \citep{wijers2019,bradley2022,tuominen2023}. Observationally estimating covering fractions of OVII in extragalactic halos is not yet possible, since this requires X-ray spectroscopy beyond the sensitivity and spectral resolution of currently available technology \citep[e.g.][]{willaims2013}. However, observations of OVII absorption from the Milky Way halo and the local group suggest that OVII is present ubiquitously out to large distances \citep{bregman2007}.

Throughout most of the halo, the convergence of OVII covering fraction with resolution is excellent, and consistent with the results from the TNG50 simulations run with slightly coarser resolutions \citep{nelson2018b}. Overall, the hot phase of the CGM is thus the best resolved, modulo possible small scale turbulence, at typical resolution levels already realised by present day cosmological simulations.

\subsection{Small Scale Structure and Cold Clouds}

\begin{figure*}
\centering 
\hspace{0.7cm}
\includegraphics[width=16cm]{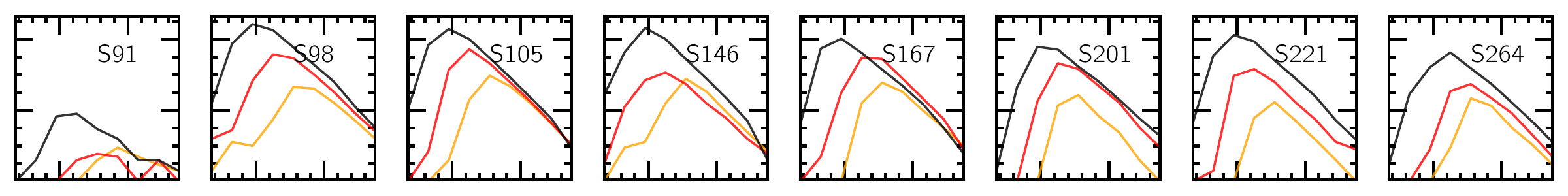}
\includegraphics[width=17.cm]{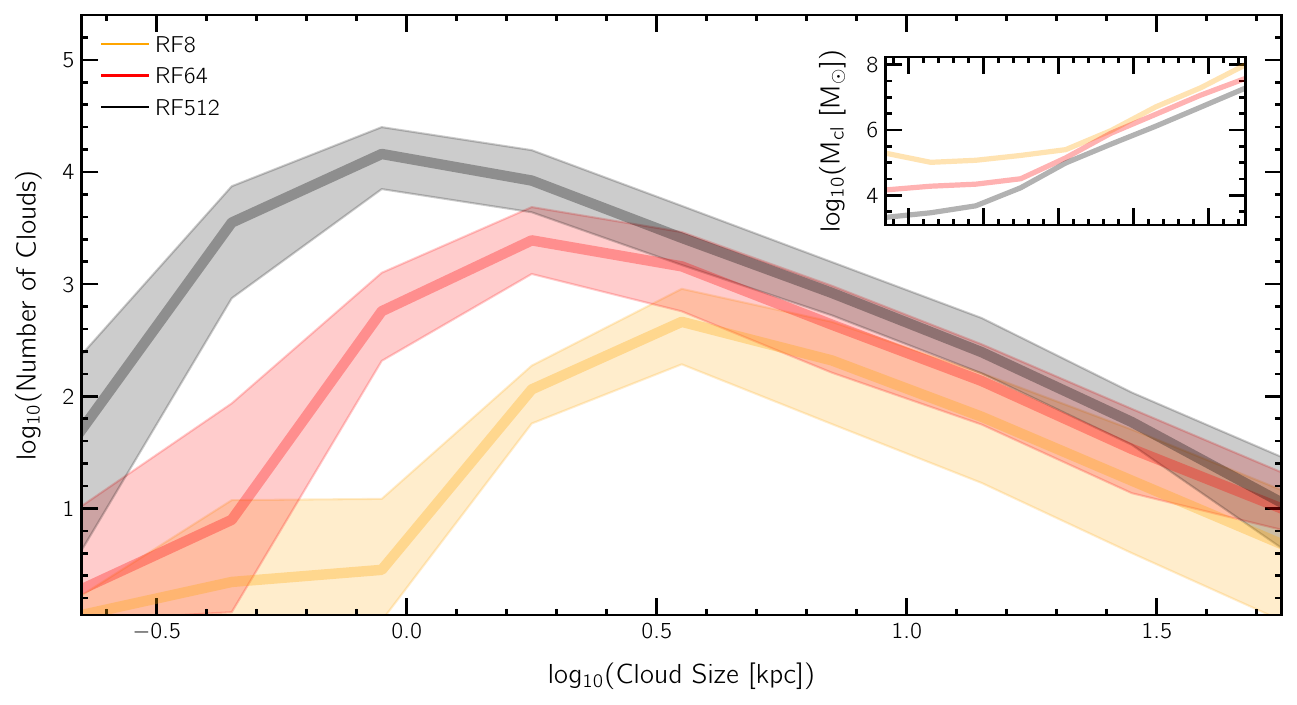}
\caption{Number of clouds as function of their size. In the main panel, medians of the three resolution runs are shown with different colors, and $16^{\rm{th}}$-$84^{\rm{th}}$ percentile values with shaded regions. Similar to Figure~\ref{fig:cloudMass}, smaller clouds are present at higher resolution, which are also greater in number. However, unlike Figure~\ref{fig:cloudMass}, the number of high-size clouds is not typically converged. We explore a possible reason for this in the inset, which shows cloud mass as a function of cloud size. The panels in the top row show the same results independently for the eight GIBLE halos.}
\label{fig:cloudSize}
\end{figure*}

We next explore the impact of numerical resolution on small scale structure in the CGM. We begin with an illustration in Figure~\ref{fig:mainVisFigureZoom}. This is similar in layout to Figure~\ref{fig:mainVisFigure}, but for a zoomed-in region of the CGM of one halo, at four different resolutions: clockwise, the resolution increases from the top left to lower left. Color indicates the local gas spatial resolution (cell radius), ranging from $\sim$\,$30$~pc in the brightest pixels, to $\sim$\,$3$~kpc in the darkest.

In the best resolved simulation (RF512; lower left), structures are clearly apparent at scales of $\sim$\,$O(1~\rm{kpc})$, with the underlying gas distribution sampled by resolution elements that are typically an order of magnitude smaller in comparison, with $\sim$\,$O(100~\rm{pc})$. Features become fuzzy as resolution decreases, with small scale structure from the RF512 run visibly absent at TNG50-2 resolution (top left panel). While the small scale structure visible in higher resolution simulations is directly due to the presence of a greater number of resolution elements, it is also possible that condensation of hot halo gas into cold, and thus denser, gas is also more efficient at increased resolution \citep[e.g.][]{joung2012}.

To quantify this small scale structure we measure the abundance of clouds that are formed by cold, dense gas. We define clouds as contiguous sets of cold gas cells ($T$\,$<$\,$10^{4.5}$~K; \citealt{ramesh2023b}), and consider all clouds down to the resolution limit, i.e. the
minimum number of cells per cloud is set to one. Owing to the nature of the Voronoi tessellation which represents our gas distribution, these clouds can trace gas structures of arbitrary sizes, masses, and shapes.

In Figure~\ref{fig:cloudMass}, we begin by exploring the distribution of cloud masses. The main panel shows the number of clouds as a function of total mass of the cloud. Curves show the median of the sample, while the shaded regions show the $16^{\rm{th}}$-$84^{\rm{th}}$ percentile values. Results from RF8, RF64 and RF512 are shown in orange, red and black, respectively. For comparison, we show results from \cite{ramesh2023b} in the dashed blue curve, i.e. the results for the TNG50 MW-like sample, from which the eight GIBLE halos were selected.

The RF8 (orange) curve peaks at roughly $10^{5.2}$~M$_\odot$, i.e. close to the resolution limit to the of the simulation, with $\sim$\,$300$ clouds on average in this bin. Clouds grow less frequent with increasing mass, with the average number dropping to $\sim$\,$30$ (4) clouds for clouds with masses $10^{7}$ ($10^{8}$)~M$_\odot$. The curve drops sharply towards the left of the peak, indicating the resolution limit. Smaller structures are simply absent beyond this point, at this resolution. The TNG50 dashed blue curve is largely similar to RF8, except with the peak shifted by $\sim$\,$0.2$~dex to the left, as a result of marginally better resolution in TNG50 with respect to RF8 (see Figure~\ref{fig:mainQuanFigure}).

The RF64 (red) curve is qualitatively similar to RF8, but with the peak shifted by a factor $\sim$\,8 to the left, signifying a better mass resolution by a factor of 8. Similarly, the RF512 (black) curve is offset horizontally further by a factor of 8. The number of `small' clouds thus increases continually with improving resolution, as signified by the arrow towards the top left of the panel. On the high mass end, very good convergence is seen between the different resolution runs. This is important since it suggests that the smallest, marginally resolved clouds present at a given resolution level are not artifacts of limited resolution, but are rather true, real structures. It is likely that these small clouds would also be present in higher resolution runs, albeit better resolved.

The eight panels in the top row show the number of clouds as a function of cloud mass for each of the eight halos separately. In each panel, very good convergence is seen between the different resolution runs at the high mass end, as in the main panel, although the first case (S91) is visibly noisy as a result of an abnormally small number of clouds.

\begin{figure}
\centering
\hspace{0.7cm}
\includegraphics[width=0.42\textwidth]{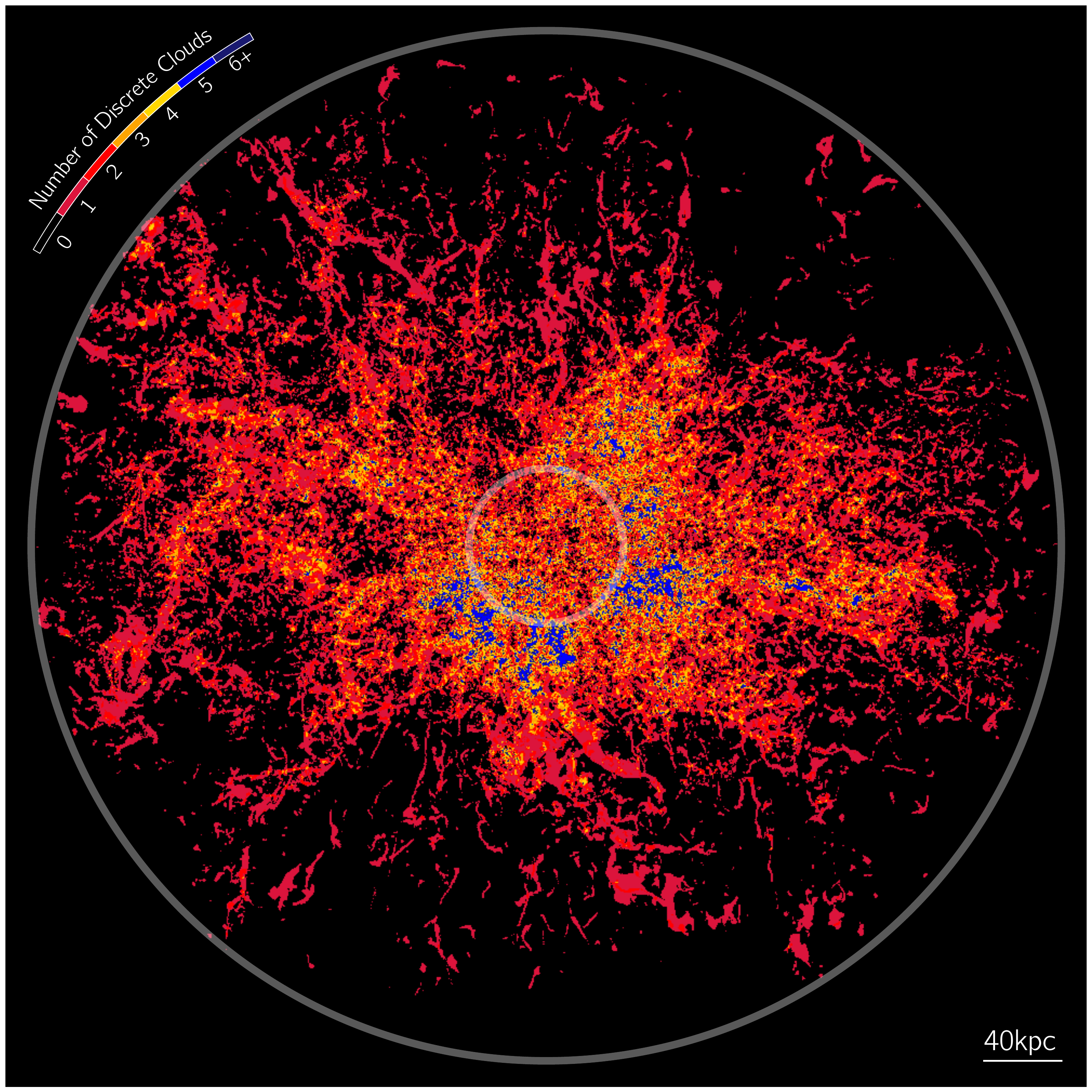}
\includegraphics[width=0.49\textwidth]{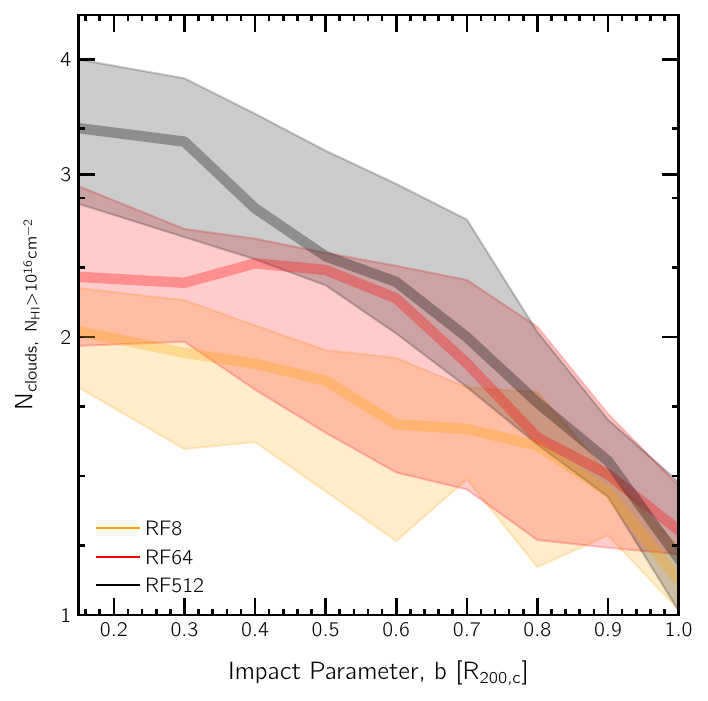}
\caption{Number of distinct (cold i.e. HI) clouds along sightlines passing through the CGM. As detailed in the main text, a cloud is counted for a given sightline if it contributes N$_{\rm{HI}}$\,$>$\,$10^{16}$\,cm$^{-2}$. The top panel shows a visualisation for the halo that contains the most clouds in our sample, with the galaxy oriented edge-on. The bottom panel shows the mean number of clouds per sightline as a function impact parameter, for those sightlines that contain at least one cloud (i.e. detectable absorption). In the innermost regions of the halo, higher resolution runs predict larger number of distinct clouds per sightline, while results are fairly well converged at larger impact parameters.}
\label{fig:coveringNumClouds}
\end{figure}

In Figure~\ref{fig:cloudSize}, we explore the distribution of cloud sizes, i.e. a measure for the spatial distribution of gas within clouds. Following \cite{ramesh2023b}, we fit the vertices of the Voronoi cells of the outer layer of each cloud to an ellipsoid, and define the size to be the mean of the lengths of the three axes of the ellipsoid. Similar to Figure~\ref{fig:cloudMass}, the main panel shows the median results from RF8, RF64 and RF512 in orange, red and black, respectively. The $16^{\rm{th}}$-$84^{\rm{th}}$ percentiles of the sample are shown with shaded bands.

The RF8 curve peaks at a cloud size of roughly $3$~kpc, with $\sim$\,$300$ clouds in this bin. The number of more massive clouds is smaller, with only $\sim$\,$10$ clouds in the high-size end ($\sim$\,$50$~kpc). As with the distribution of cloud masses, the curve drops rapidly to the left of the peak, signifying the resolution limit of the simulation. The RF64 curve peaks at $\sim$\,$1.5$~kpc, i.e. roughly a factor of two smaller than RF8, corresponding to a factor of eight in mass (or resolution elements). Similarly, the RF512 curve is further offset by a factor of $\sim$\,2, peaking at $\sim$\,$800$~pc. As before, with increasing resolution, the smallest objects are always more in number. 

However, unlike the case of cloud masses, the number of high-size objects does not converge as rapidly, as can be seen by the vertical offsets between different curves at the right side of the main panel. To understand this trend, in the inset on the top-right, we show the relation between cloud masses (y-axis) and cloud size (x-axis). Note that the range of the x-axis of the inset, and the position of ticks, is exactly the same as that of the x-axis of the main panel. For visual clarity, we avoid showing the percentile regions, and only plot the medians in the inset.

At fixed cloud size, on average, cloud masses are lower at better resolution levels. For instance, at a size of $10$~kpc, clouds have an average mass of $\sim$\,$10^{6.5}$~M$_\odot$ in the RF8 simulations. At RF64-resolution, these clouds are slightly less massive at $\sim$\,$10^{6.35}$~M$_\odot$. At even higher resolution, analogous clouds in the RF512 simulations have an average mass of $\sim$\,$10^{6}$~M$_\odot$. At lower cloud sizes, there are intermediate regions where the RF8 median seems to converge with RF64 (around a cloud size of $6$~kpc), and RF64 with RF512 (around a size of $3$~kpc). However, this is likely an effect of clouds in the lower resolution simulation being poorly resolved, i.e. these `low-mass'/`small-size' clouds typically are composed of but a handful of Voronoi cells. We interpret this decrease in mass (at fixed cloud size) to be the result of their spatial distribution being captured to a better extent by the increased number of resolution elements \citep[see also][]{hummels2019}. This could also be a sign of clouds, at low resolution, having artificially low pressure, such that they need a greater amount of mass for a given size to achieve pressure equilibrium, although this needs to be analyzed in more detail.

This explains the lack of convergence between the medians of the three resolution levels at the high-size end: at fixed size, cloud masses are smaller at better resolution, and these clouds are typically more in number (Figure~\ref{fig:cloudMass}). As a result, the higher resolution curves are shifted vertically upwards and horizontally towards the right, leading to a slower apparent convergence. However, for clouds above a certain mass threshold, we suspect sizes to be converged across resolutions, since such massive clouds are always resolved by a large number of resolution elements. This evolution towards convergence can be seen at the right-most edge of the main panel, where the RF64 median asymptotically approaches that of RF512. Unfortunately, the most massive clouds available with the GIBLE sample are not sufficient to demonstrate this for the RF8 runs. Comparison with a future, even higher resolution simulation is required to test this hypothesis.

Similar to Figure~\ref{fig:cloudMass}, the panels in the top row show the number of clouds as a function of cloud size, for the eight GIBLE halos separately. While most cases are in agreement with the conclusions of the main panel, many actually show decent convergence between the three resolution levels at the high-size end. This suggests that the lack of convergence clearly seen in the last two halos, but not at all in the third, fourth, or fifth may be due to the particular state or evolution of the CGM of those halos.

\subsection{Cloud Abundances along Absorption Sightlines}

While the earliest observations of (high velocity) clouds in the Milky Way sky were made possible through HI emission studies \citep{muller1963,wakker1991,wakker1997}, more recent explorations have begun using absorption line measurements to identify cloud like features in the Milky Way \citep[e.g.][]{lehner2012,richter2017} and extragalactic halos \citep[e.g.][]{zahedy2019,weng2022}. In the latter case, the number of clouds along a given line of sight can be estimated through the number of Voigt profile components required to fit the observed spectrum.

In Figure~\ref{fig:coveringNumClouds}, we study the impact that numerical resolution has on making predictions for observations of this kind. For simplicity, we do not make synthetic spectra to exactly mimic observational studies, but rather adopt a simpler approach possible with the simulations. Specifically, we project the mass of each cloud onto a grid of pixels using the standard SPH-kernel, and for every pixel that the cloud contributes a HI column density $>$\,$10^{16}$\,cm$^{-2}$ to, the cloud count of that pixel is increased by one. Each contribution of N$_{\rm{HI}}$\,$>$\,$10^{16}$\,cm$^{-2}$ by a cloud is thus assumed to create a measure-able dip in an absorption spectrum. Further, it is assumed that every cloud produces a distinct dip in the spectrum, something which may not always be true, for e.g. if two clouds along the line of sight have very similar velocities. Further, the impact of limited spectral resolution and/or presence of noise is ignored. What follows is thus meant as a zeroth-order theoretical study for the number of discrete clouds along a line of sight through the CGM.

The top panel visually demonstrates this technique. We choose the halo with the most clouds for this purpose (S98, RF512), with the central galaxy oriented edge-on at the center. The image extends $\pm$\,$1.05~R_{\rm{200,c}}$ from edge to edge along the plane of the image, and $\pm$\,$1~R_{\rm{200,c}}$ in the perpendicular direction. We draw two circles at $[0.15, 1.00]~R_{\rm{200,c}}$, signifying the adopted boundaries of the CGM. Only clouds between these two (3D) boundaries are considered for this analysis, even though clouds are visible within the inner circle as a result of projection effects.

Colors show the number of (HI i.e. cold) clouds along each sightline. Sightlines with no clouds are colored black, while those with only one cloud are dark red, two clouds are light red, three clouds are orange, four clouds are gold, five clouds are blue, and six or more clouds are midnight blue, as shown by the discrete colorbar shown on the top-left of the panel. There is a visible radial trend with more clouds per sightline, on average, at smaller impact parameters, with typical colors shifting from gold/blue close to $0.15~R_{\rm{200,c}}$ (4+ clouds per sightline) to dark red/red close to the virial radius ($1-2$ clouds). This is broadly consistent with our earlier findings from the TNG50 simulation, where a larger number of clouds are seen at smaller galactocentric distances, each of which are smaller than their most distant counterparts \citep{ramesh2023b}. However, even at fixed impact parameter, there is a large variation: for instance, in the innermost regions of the halo, there are clearly sightlines with $1-2$ clouds adjacent to other sightlines with $4-6+$ clouds. Gravitational lens absorption studies, i.e. those that probe multiple sightlines of a single foreground CGM using different (lensed) images of the background source \citep[e.g.][]{augustin2021}, should expect to see significant
variability in component structure between different sightlines.

In the lower panel, we quantify the radial trend discussed above. For a given impact parameter, we compute the mean number of clouds along those sightlines that contain \textit{at least} one cloud. This metric is thus \text{it} not a measure of the covering fraction of clouds, a related quantity which has already been explored earlier (Figure~\ref{fig:coveringFracHI}), but rather describes the expected number of cold clouds along those sightlines for which an absorption signal is present. For each galaxy, we construct radial profiles for $100$ random orientations, and compute their median. The three solid curves show the median of these radial profiles for the sample, while the shaded regions show the $16^{\rm{th}}$-$84^{\rm{th}}$ percentiles variations. Results corresponding to RF8, RF64 and RF512 are shown in orange, red and black, respectively.

In our best resolved simulations (RF512), at an impact parameter of 
$\sim$\,$0.15~R_{\rm{200,c}}$, sightlines that contain clouds have multiple ($3-4$), on average. The number steadily drops to $2-3$ at $\sim$\,$0.5~R_{\rm{200,c}}$, and further to $\sim$\,$1$ at the virial radius. At RF64 resolution, clouds are fewer in number at impact parameters $\lesssim$\,$0.5~R_{\rm{200,c}}$ ($2-3$ clouds), but numbers are converged with RF512 at larger distances. At an even coarser resolution of RF8, numbers are typically smaller out to even larger distances ($\sim$\,$0.8~R_{\rm{200,c}}$), beyond which decent convergence is seen with higher resolution runs. We suspect that this convergence at large impact parameters is a result of clouds, on average, being more massive at larger distances \citep{ramesh2023b}, and hence are typically better resolved across different resolution levels (Figure~\ref{fig:cloudMass}). Overall, our RF512 runs suggest that absorption spectra of cold species at low impact parameters will have complex spectral morphologies, sometimes composed of a half dozen or more discrete absorbing clouds.

\subsection{Small-scale density and velocity structure}

\begin{figure*}
\centering 
\includegraphics[width=8.cm]{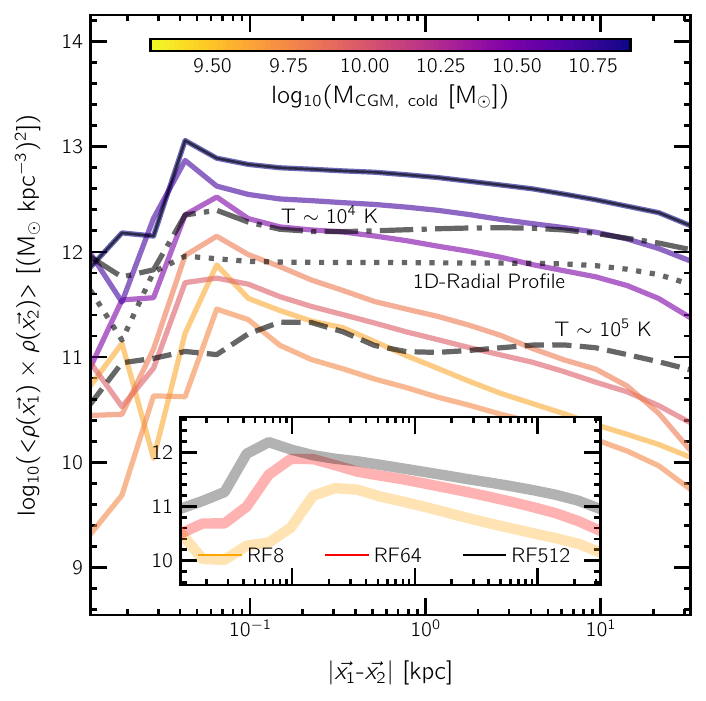}
\includegraphics[width=8.06cm]{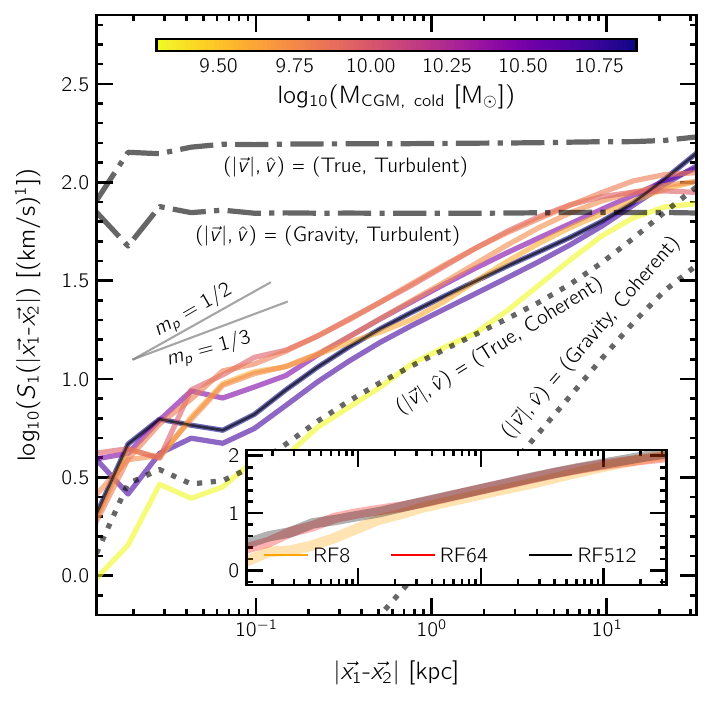}
\caption{Quantifying small scale structure of CGM gas: the left panel shows the auto-correlation function of the density field, while the right panel shows the first-order velocity structure function. The main panels show results for the eight halos from the RF512 run, with curves colored by the cold gas mass in the CGM of the corresponding halo. Insets assess numerical convergence by comparing the median results of the RF8 (orange), RF64 (red) and RF512 (black) runs. The dashed, dot-dashed and dotted curves correspond to a variety of test cases included for comparison, as elaborated in the main text.}
\label{fig:smallScale}
\end{figure*}

In the final part of this paper, we explore two additional metrics to quantify the small-scale structure of CGM gas. In Figure~\ref{fig:smallScale}, we show the auto-correlation function of the density field (left) and the first-order velocity structure function (VSF; right). Both these metrics describe statistical quantities related to pairs of gas cells at positions $\vec{x_1}$ and $\vec{x_2}$, thus separated by distance $|\vec{x_1} - \vec{x_2}|$: while the former shows how the densities at these points are related ($<\rho(\vec{x_1}) \times \rho(\vec{x_2})>$), the latter describes the differences in velocities ($S_1(|\vec{x_1} - \vec{x_2}|) = <|\vec{v}(\vec{x_1}) - \vec{v}(\vec{x_2})|>$).\footnote{We use the positions of simulated gas cells themselves as tracers, and given that all (CGM) gas cells have roughly the same mass (Section~\ref{methods}), these statistics are therefore essentially mass-weighted. An alternate approach would be to sample points randomly (or evenly) throughout the halo, providing a more volume-weighted result, although we do not explore this approach here.}

In the main panels, we show results of the eight halos from the RF512 runs, with each solid curve colored according to the total CGM cold gas mass of that halo at $z=0$. Additionally, for one halo (indicated with the thin overlaid black curve), we show a variety of test cases, further described below. The insets in each panel shown the medians of the eight halos, for the RF8, RF64 and RF512 runs separately, in orange, red and black, respectively. 

In the main panel on the left, curves peak close to $40-60$~pc, corresponding to two times the best spatial resolution achieved in the CGM (Figure~\ref{fig:mainQuanFigure}). This peak, and the sharp drop to the left, is an artifact of finite resolution, and represents the resolution limit of the simulations. To the right of the peak, each curve decays with a finite slope. The amplitude and location of the peak, as well as the slope of the auto-correlation function at larger scales, all depend on the total CGM cold gas mass.

To unravel some of the drivers of these trends, we conduct a number of tests. The dotted black line shows the auto-correlation function for a case where every gas cell is assigned a new density, given by the one-dimensional radial profile of the halo. This curve thus assesses the impact of the radial dependence of density on the auto-correlation function. Similar to the real curves, this test case drops sharply to the left of $\sim 40$~pc, again signifying the resolution limit. However, at larger separations, the curve is more or less flat, at least out to $\sim 10-20$~kpc. The finite slope seen in the real case is thus not an effect of the radial profile alone.

In the black dot-dashed and dashed curves, we show results when only T~$\sim$~$10^4$~K and $10^5$~K gas is included, respectively, with the latter vertically offset by 2~dex for better visibility. Similar to the dotted curve, these are flat out to $\sim 10-20$~kpc. However, they have different amplitudes, corresponding to different mean densities at those temperatures, and they peak at different values, tracing the density dependence of spatial resolution. We conclude that the slopes of the actual curves, i.e. the decrease of the density auto-correlation function with increasing spatial scale, arises due to an averaging across a `fan' of contributing components, corresponding to different temperatures/densities, each of which have a different amplitude, and each of which dominates at different characteristic separations. This also explains the dependence of the slope on the cold gas content of the CGM: in halos with less cold gas, the warm-hot phase dominates the auto-correlation at smaller distances to a larger extent, giving rise to a steeper slope.

The inset shows the impact of numerical resolution. With improving resolution, the small-scale peak shifts towards smaller separations, as expected. Additionally, the amplitude of the auto-correlation function increases, signifying the presence of more dense gas at improved resolutions. The slopes of these median curves to the right of the peak are roughly the same irrespective of resolution, indicating that the fraction of gas in different phases is typically not a strong function of resolution (Table~\ref{table:halo_summary}). 

For comparison, analogous metrics are used to quantify the level of clustering of sun-spots on the solar surface \citep[e.g.][]{zhou2020}, star clusters in galaxies \citep[e.g.][]{grasha2017}, and galaxies in the universe \citep[e.g.][]{keihanen2019}. In all cases, a given finite slope corresponds to greater clustering at smaller scales. Our measurement is also closely related to other metrics of cloud-cloud clustering, such as the $\Delta_{10}$ measurement used in \cite{ramesh2023b}. This phase-dependent (i.e. temperature binned) density auto-correlation could be a valuable tool to quantify the cold-phase vs warmer-phase of the CGM in simulations, and possibly also a way to, for e.g., compare different simulations with different physics and across different numerical resolutions.

The right panel of Figure~\ref{fig:smallScale} moves from density to velocity, showing the first-order velocity structure function of CGM gas from $\sim 10$\,pc to $\sim 40$\,kpc scales. All curves steadily increase with increasing separation, with a typical slope of $m_p$\,$\sim$\,$0.45$ (VSF($r) \propto r^{m_p}$), before plateauing at larger separations (not explicitly shown). This slope is steeper than that predicted by Kolmogorov turbulence ($m_p$\,$=$\,$1/3$; \citealt{kolmogorov1941}), and slightly smaller than a value of $m_p$\,$=$\,$1/2$ \citep{burgers1948}, both of which are shown by light gray curves in the main panel. Pairs of gas cells closer to each other are thus more likely to be moving with similar velocities. Similar to the left panel, a trend is present with respect to CGM cold gas mass, albeit not as pronounced. The inset assesses numerical convergence between the RF-runs: at small separations ($\lesssim$\,$200$~pc), RF64 and RF512 yield similar results, with RF8 under-predicting values, although all three resolution runs are converged at larger separations.

In the various dotted and dot-dashed lines, we compare with test cases where the motion of gas is prescribed to be fully turbulent or coherent/laminar. To do so, we first consider every gas cell to be on a perfectly circular orbit about the centre of the halo. For a gas cell at position $(x,y,z)$ and total speed $v$, this would imply

\begin{equation}
    x v_x + y v_y + z v_z = 0; \,\,\,
    v_x^2 + v_y^2 + v_z^2 = v^2.
\end{equation}

where $(v_x, v_y, v_z)$ are the velocity components along the three different axes. With three variables (i.e. the three velocity components) and only two equations, we adopt $v_z$ as a necessarily free parameter. To ensure that the solution to the above set of equations is real, values of $v_z$ are constrained in the interval $[-v_{z,max}, +v_{z,max}]$, where $v_{z,max}$ is a function of $(x, y, z, v)$. Setting $v_z$ to $0$ for all gas cells yields a solution where the flow is coherent, or at least as coherent as it can be for this spherical-flow assumption, while randomly varying $v_z$ between $\pm v_{z,max}$ results in a flow that is turbulent.

We proceed to set the values of $v$ in a number of different ways. First, we assume gravity-only motion i.e. where the total speed of each gas cell is set such that the centrifugal force perfectly balances the gravitational force of all particles enclosed within its orbit. In addition, we consider the case where the speed of each gas cell is taken as is from the simulation, i.e. in this case, the magnitude of the velocity vector is `true', but the direction is altered. For each of these two $v$-cases, we show the resulting velocity structure function for the coherent (dotted) and fully turbulent (dot-dashed) cases.

The fully turbulent cases do not contain much information, with the curves showing almost no trend with scale, except for noise at small separations as a result of limited resolution. As expected, the gravity-only cases plateau at smaller values in comparison to the true case, showing that speeds of gas cells are impacted by various hydrodynamical interactions, in addition to gravity.

The coherent cases (dotted curves) are more informative. Both cases show an increasing slope towards larger separations. The slope for the gravity-only case is larger in comparison to the `true' values test case, although the actual VSF values are smaller at all distances. As with the above discussion, this too shows that speeds of gas cells are impacted by various hydrodynamical interactions. Interestingly, the slope of the true-velocity case is similar to the real case, but the two curves are offset by $\sim$\,$0.3$~dex. CGM gas in Milky Way-like halos is thus clearly neither in coherent spherical motion nor fully turbulent motion.

For comparison, various studies have explored the velocity structure function of halo gas, albeit typically with idealised simulations where resolutions can be higher. For instance, \cite{hillel2020} analyse simulations of intracluster medium (ICM) gas being heated by jet-inflated hot bubbles. They find that, depending on the seperation scale, slopes can either be steeper or shallower than $m_p$\,$=$\,$1/3$, i.e. Kolmogorov turbulence. \cite{wang2021} study the impact of AGN jets in a Perseus-like cluster, using both hydrodynamical (HD) and magnetohydrodynamical (MHD) simulations. For the hot phase of gas, they find slightly shallower slopes in the MHD case ($m_p$\,$\lesssim$\,$0.5$) than the HD case ($m_p$\,$\sim$\,$0.5$), with a similar trend for the cold phase as well.

Recent observations have studied velocity structure functions in a number of nearby clusters. For instance, \cite{li2020} probe cool gas at the centres of clusters through their H$\alpha$ emission, and find a slope of $m_p$\,$\sim$\,$0.5$ for Persesus, and steeper slopes of $m_p$\,$\sim$\,$0.6$  for Abell 2597 and for the inner $2.5$~kpc of the Virgo cluster. Values of slopes can clearly vary a lot, signifying varied levels of turbulence over different scales and environments. Although the various studies mentioned here focus on very different environments than that of halos like the Milky Way, we mention them here to provide a sense of typical values.
Moreover, note that the first two studies mentioned above, as well as the results from GIBLE, use a three-dimensional VSF, in contrast to observations like \cite{li2020} that are limited to a projection-based two-dimensional VSF.

The velocity structure function is clearly a useful tool to gain some understanding of, and quantify, the kinematic structure of CGM gas. Similar to the density auto-correlation, this could be a powerful method to compare simulations run with different models and/or resolutions. Moreover, one could compute a two-dimensional projected VSF to mock observations studies, thereby improving the theoretical understanding and the interpretation of such results \citep[e.g.][]{mohapatra2022}. 
Future X-ray
spectroscopic imaging concept missions such as the Line Emission Mapper
(LEM; \citealt{kraft2022}) will enable us to measure the VSF of the hot CGM for
Milky Way like galaxies in the future.
Comparisons with such observations would likely prove useful to test and validate different feedback models in simulations, which could have an impact of the VSF of CGM gas.

\section{Summary and Conclusions}\label{summary}

In this paper we introduce the GIBLE simulations, a new suite of cosmological hydrodynamical zoom-in simulations of galaxies with a `CGM refinement' method, designed to increase numerical resolution in the circumgalactic medium. While gas in the galaxy is maintained at a lower resolution, CGM gas is super-refined to significantly better mass (and spatial) resolution. Here we introduce a sample of eight, Milky Way-like galaxies, drawn originally from the TNG50 simulation, which reach a mass resolution of $\sim$\,$10^3\, \rm{M_\odot}$ in the CGM, competitive with current state-of-the-art zoom-in simulations of Milky Way-like galaxies at $z=0$. The high computational efficiency of our CGM refinement method enables us to simulate a relatively large and diverse set of halos in order to explore their diversity at high resolution. Our main findings of this presentation paper, the first in the GIBLE series, are:

\begin{enumerate}

    \item Increased CGM gas resolution not only results in the halo being more finely resolved in general, but also in visibly apparent structure on smaller scales (Figure~\ref{fig:mainVisFigure}). In our best resolved runs, we achieve a median spatial resolution of $\sim$\,$75$~pc at the inner boundary of the CGM ($0.15~R_{\rm{200,c}}$), which coarsens with increasing distance to $\sim$\,$700$~pc at the virial radius. The cold phase of gas, however, is better spatially resolved at all distances (Figure~\ref{fig:mainQuanFigure}).

    \item Integrated (i.e. global) properties of the halo that we explore are well converged at the resolutions we achieve (Figure~\ref{fig:table_sum} and Table~\ref{table:halo_summary}). These include the CGM gas mass, cold gas fraction, temperature, and volume fractions of cold, warm, and hot phases. Time-averaged, sample-averaged PDFs of gas properties -- temperature, metallicity, magnetic field strength, and velocity -- are also well converged (Figure~\ref{fig:nonVaryingHist}). The notable exception is the abundance of cold CGM gas clouds, the number of which continues to increase with increasing resolution.

    \item While some key observables are well converged with resolution, others are not. For instance, the HI covering fraction of pLLS systems (N$_{\rm{HI}}$\,$>$\,$10^{16}$\,cm$^{-2}$) is only converged at resolutions $\lesssim$\,$10^4$\,$\rm{M_\odot}$, and simulations run at a resolution of $\sim$\,$10^5$\,$\rm{M_\odot}$ under predict covering fractions out to a distance of $\sim$\,$0.8$\,$R_{\rm{200,c}}$. Covering fractions of more dense gas, i.e. sub-DLA and DLA systems, are not converged at our resolutions (Figure~\ref{fig:coveringFracHI}).

    \item Similar trends are also observed for the covering fraction of MgII and CIV, which are tracers of the cold- and warm-phase of gas, respectively. On the other hand, the diffuse, volume-filling hot phase traced by OVII is well converged at these resolutions (Figure~\ref{fig:coveringFrac}).

    \item As expected, better resolved simulations produce structures on smaller scales (Figure~\ref{fig:mainVisFigureZoom}). One way to quantify this is by characterizing the number and properties of cold, dense gas clouds. At better resolutions, the CGM has a larger number of such clouds, with the number increasing steeply as a power law. Importantly, the number of massive clouds, i.e. clouds well above the resolution limit, are converged across resolution levels (Figure~\ref{fig:cloudMass}).

    \item This has an impact on predictions for, e.g., the number of pLLS absorbers along a given line of sight through the CGM. Better resolved simulations predict a larger number of absorbers at small impact parameters, although results approach convergence at larger impact parameters, where clouds are bigger. While the median number of absorbers in our highest resolution runs varies between $1$ and $4$, depending on impact parameter, absorption spectra along some sightlines may have more complex spectral morphologies, with possibly $6$ or more distinct absorbers (Figure~\ref{fig:coveringNumClouds}).

    \item We also quantify small-scale structure through the auto-correlation of the density field and velocity structure function. Both metrics yield valuable information about features in CGM gas. The former shows a finite slope with the auto-correlation decreasing towards larger separations, signifying that smaller-separation pairs are typically dominated by colder, denser gas. The latter shows that CGM gas is not completely turbulent, and not governed by gravity alone, but rather by added hydrodynamic interactions, with values close to converged at resolutions $\lesssim$\,$10^4$\,$\rm{M_\odot}$ (Figure~\ref{fig:smallScale}).
    
\end{enumerate}

Project GIBLE aims to bridge the gap between cosmological and idealised simulations. However, despite achieving a resolution close to the (currently) best resolved cosmological zoom-in simulation of a Milky Way-like galaxy, we are still unable to resolve parsec-scale features in the CGM. Future extensions of GIBLE will aim to push the resolution significantly higher than this first series of simulations. Further, the current simulations omit
potentially important physical processes such as cosmic rays or thermal conduction. Future simulations will aim to tackle these shortcomings.

\section*{Data Availability}

Collaboration on new projects with the GIBLE simulation data is welcome, and potential collaborators are encouraged to contact the authors directly. Data related to this publication is available upon reasonable request to the corresponding author. 

\section*{Acknowledgements}

RR and DN acknowledge funding from the Deutsche Forschungsgemeinschaft (DFG) through an Emmy Noether Research Group (grant number NE 2441/1-1). RR is a Fellow of the International Max Planck Research School for Astronomy and Cosmic Physics at the University of Heidelberg (IMPRS-HD). The authors thank Volker Springel for providing access to the private branch of the \textsc{arepo} code, Annalisa Pillepich for a helpful discussion on selecting a diverse sample of Milky Way-like galaxies, Christoph Federrath for suggestions on the velocity structure function analysis, and Cameron Hummels, Freeke van de Voort and Robert J J Grand for sharing relevant data points. This work has made use of the VERA supercomputer of the Max Planck Institute for Astronomy (MPIA), the COBRA supercomputer, both operated by the Max Planck Computational Data Facility (MPCDF), and the HELIX supercomputer of the University of Heidelberg. The authors acknowledge support by the state of Baden-Württemberg through bwHPC and the German Research Foundation (DFG) through grant INST 35/1597-1 FUGG. This research has made use of NASA's Astrophysics Data System Bibliographic Services. 

\bibliographystyle{mnras}
\bibliography{references}

\end{document}